\documentclass[aps,prl,reprint,superscriptaddress,longbibliography,floatfix]{revtex4-2}

\usepackage{amsmath,amssymb,bm,mathtools}
\usepackage{graphicx}
\usepackage{xcolor}
\usepackage{etoolbox}
\usepackage{microtype}

\usepackage[colorlinks=true,allcolors=blue]{hyperref}

\definecolor{revisiongreen}{RGB}{0,125,45}

\newcommand{\Tr}{\operatorname{Tr}}
\newcommand{\id}{\operatorname{id}}
\newcommand{\cD}{\mathcal D}
\newcommand{\cJ}{\mathcal J}
\newcommand{\ket}[1]{|#1\rangle}
\newcommand{\bra}[1]{\langle#1|}

\begin{document}
\title{Force Sensing in a Recoverable Bosonic Quantum Memory}
\author{Tairong Bai}
\thanks{These authors contributed equally to this work.}
\affiliation{National Laboratory of Solid State Microstructures, School of Physics, Nanjing University, Nanjing 210093, China}
\affiliation{Shishan Laboratory, Nanjing University, Suzhou 215163, China}
\affiliation{Jiangsu Key Laboratory of Quantum Information Science and Technology, Nanjing University, Suzhou 215163, China}
\author{Wen Zheng}
\thanks{These authors contributed equally to this work}
\email{zhengwen@nju.edu.cn}
\affiliation{National Laboratory of Solid State Microstructures, School of Physics, Nanjing University, Nanjing 210093, China}
\affiliation{Shishan Laboratory, Nanjing University, Suzhou 215163, China}
\affiliation{Jiangsu Key Laboratory of Quantum Information Science and Technology, Nanjing University, Suzhou 215163, China}
\affiliation{ LaunchQ (Suzhou) Technology Co., Ltd., Suzhou 215163, China}
\author{Haoyang Cai}
\affiliation{National Laboratory of Solid State Microstructures, School of Physics, Nanjing University, Nanjing 210093, China}
\affiliation{Shishan Laboratory, Nanjing University, Suzhou 215163, China}
\affiliation{Jiangsu Key Laboratory of Quantum Information Science and Technology, Nanjing University, Suzhou 215163, China}
\affiliation{Synergetic Innovation Center of Quantum Information and Quantum Physics, University of Science and Technology of China, Hefei, Anhui 230026, China}
\affiliation{Hefei National Laboratory, Hefei 230088, China}

% \author{Xiang Deng}
% \affiliation{National Laboratory of Solid State Microstructures, School of Physics, Nanjing University, Nanjing 210093, China}
% \affiliation{Shishan Laboratory, Nanjing University, Suzhou 215163, China}
% \affiliation{Jiangsu Key Laboratory of Quantum Information Science and Technology, Nanjing University, Suzhou 215163, China}
% \author{Haoyu Zhou}
% \affiliation{National Laboratory of Solid State Microstructures, School of Physics, Nanjing University, Nanjing 210093, China}
% \affiliation{Shishan Laboratory, Nanjing University, Suzhou 215163, China}
% \affiliation{Jiangsu Key Laboratory of Quantum Information Science and Technology, Nanjing University, Suzhou 215163, China}
% \author{Tao Zhang}
% \affiliation{National Laboratory of Solid State Microstructures, School of Physics, Nanjing University, Nanjing 210093, China}
% \affiliation{Shishan Laboratory, Nanjing University, Suzhou 215163, China}
% \affiliation{Jiangsu Key Laboratory of Quantum Information Science and Technology, Nanjing University, Suzhou 215163, China}
% \author{Xudong Liao}
% \affiliation{National Laboratory of Solid State Microstructures, School of Physics, Nanjing University, Nanjing 210093, China}
% \affiliation{Shishan Laboratory, Nanjing University, Suzhou 215163, China}
% \affiliation{Jiangsu Key Laboratory of Quantum Information Science and Technology, Nanjing University, Suzhou 215163, China}
\author{Xianke Li}
\affiliation{National Laboratory of Solid State Microstructures, School of Physics, Nanjing University, Nanjing 210093, China}
\affiliation{Shishan Laboratory, Nanjing University, Suzhou 215163, China}
\affiliation{Jiangsu Key Laboratory of Quantum Information Science and Technology, Nanjing University, Suzhou 215163, China}

\author{Yangyang Ge}
\affiliation{National Laboratory of Solid State Microstructures, School of Physics, Nanjing University, Nanjing 210093, China}
\affiliation{Shishan Laboratory, Nanjing University, Suzhou 215163, China}
\affiliation{Jiangsu Key Laboratory of Quantum Information Science and Technology, Nanjing University, Suzhou 215163, China}

\author{Kun Zhou}
\affiliation{National Laboratory of Solid State Microstructures, School of Physics, Nanjing University, Nanjing 210093, China}
\affiliation{Shishan Laboratory, Nanjing University, Suzhou 215163, China}
\affiliation{Jiangsu Key Laboratory of Quantum Information Science and Technology, Nanjing University, Suzhou 215163, China}
\affiliation{ LaunchQ (Suzhou) Technology Co., Ltd., Suzhou 215163, China}
\author{Xiangmin Yu}
\affiliation{National Laboratory of Solid State Microstructures, School of Physics, Nanjing University, Nanjing 210093, China}
\affiliation{Shishan Laboratory, Nanjing University, Suzhou 215163, China}
\affiliation{Jiangsu Key Laboratory of Quantum Information Science and Technology, Nanjing University, Suzhou 215163, China}

\author{Yu Zhang}
\affiliation{National Laboratory of Solid State Microstructures, School of Physics, Nanjing University, Nanjing 210093, China}
\affiliation{Shishan Laboratory, Nanjing University, Suzhou 215163, China}
\affiliation{Jiangsu Key Laboratory of Quantum Information Science and Technology, Nanjing University, Suzhou 215163, China}
\author{Shaoxiong Li}
\affiliation{National Laboratory of Solid State Microstructures, School of Physics, Nanjing University, Nanjing 210093, China}
\affiliation{Shishan Laboratory, Nanjing University, Suzhou 215163, China}
\affiliation{Jiangsu Key Laboratory of Quantum Information Science and Technology, Nanjing University, Suzhou 215163, China}
\affiliation{Synergetic Innovation Center of Quantum Information and Quantum Physics, University of Science and Technology of China, Hefei, Anhui 230026, China}
\affiliation{Hefei National Laboratory, Hefei 230088, China}
\author{Xinsheng Tan}
\affiliation{National Laboratory of Solid State Microstructures, School of Physics, Nanjing University, Nanjing 210093, China}
\affiliation{Shishan Laboratory, Nanjing University, Suzhou 215163, China}
\affiliation{Jiangsu Key Laboratory of Quantum Information Science and Technology, Nanjing University, Suzhou 215163, China}
\affiliation{Synergetic Innovation Center of Quantum Information and Quantum Physics, University of Science and Technology of China, Hefei, Anhui 230026, China}
\affiliation{Hefei National Laboratory, Hefei 230088, China}

\author{Jie Zhao}
\affiliation{National Laboratory of Solid State Microstructures, School of Physics, Nanjing University, Nanjing 210093, China}
\affiliation{Shishan Laboratory, Nanjing University, Suzhou 215163, China}
\affiliation{Jiangsu Key Laboratory of Quantum Information Science and Technology, Nanjing University, Suzhou 215163, China}
\author{Yang Yu}
\email{yuyang@nju.edu.cn}
\affiliation{National Laboratory of Solid State Microstructures, School of Physics, Nanjing University, Nanjing 210093, China}
\affiliation{Shishan Laboratory, Nanjing University, Suzhou 215163, China}
\affiliation{Jiangsu Key Laboratory of Quantum Information Science and Technology, Nanjing University, Suzhou 215163, China}
\affiliation{ LaunchQ (Suzhou) Technology Co., Ltd., Suzhou 215163, China}
\affiliation{Synergetic Innovation Center of Quantum Information and Quantum Physics, University of Science and Technology of China, Hefei, Anhui 230026, China}
\affiliation{Hefei National Laboratory, Hefei 230088, China}

\date{\today}

\begin{abstract}
  %  Sensing and quantum error correction place competing demands on a bosonic mode: weak external displacements carry the signal but may also be suppressed as errors.
   In continuous-variable quantum information, sensing and quantum error correction in the same mode face an intrinsic trade-off:
   sensors must accumulate weak external displacements, whereas bosonic codes typically suppress such shifts as errors. 
   Here, we establish compatibility by constructing finite-energy bosonic codes that enable a single oscillator to sense an external force while actively protecting an unknown logical qubit.
   By encoding logical amplitudes across discrete wave-packet sites and capturing the force signal in a shared internal coordinate,
   we achieve an endpoint probe readout that is independent of the logical state under continuous photon loss $\gamma$.
   Crucially, binomial site recovery cancels packet-dependent probe distortions alongside logical errors:
   for a $(2r+2)$-packet code, memory infidelity is suppressed to $O(\gamma^{r+1})$ while residual memory-probe correlations scale as $O(\gamma^{r+2})$. 
  %  Converged physical-channel simulations confirm this hierarchy and resolve cubic memory-probe decorrelation for the four-packet code.
   Finally, superconducting quantum circuit benchmarks under competing photon-number dephasing and Kerr nonlinearity identify preferred encoding regimes for hardware implementation.
% Here, we construct finite-energy bosonic codes that enable continuous force sensing while protecting an unknown logical qubit within the same oscillator.
% Logical information is encoded in binomial amplitudes across discrete wave-packet sites, while a shared internal coordinate carries the force signal.
% Under continuous vacuum photon loss and ideal endpoint controls, the probe readout is exactly independent of the logical state.
% Binomial site recovery cancels packet-dependent probe distortions together with leading logical errors:
% for a $(2r+2)$-packet code, memory infidelity scales as $O((\kappa T)^{r+1})$, while residual memory--probe correlations are suppressed to $O((\kappa T)^{r+2})$, where $\kappa$ is the photon-loss rate and $T$ the sensing duration.
% Converged physical-channel simulations verify the memory-infidelity hierarchy and resolve cubic memory--probe decorrelation for the four-packet code.
% Superconducting-cavity benchmarks incorporating photon-number dephasing and Kerr nonlinearity reveal how competing noise mechanisms determine the preferred encoding.
\end{abstract}
\maketitle

\textcolor{blue}{\emph{Introduction.---}}%
Quantum sensing infers an external field from a system's dynamical response~\cite{degen,giovannetti},
whereas quantum error correction protects an unknown logical state and its entanglement from environmental noise~\cite{shor,bennett}.
% Can a single physical oscillator perform both tasks simultaneously?
Can an occupied oscillator perform both tasks simultaneously without exposing or sacrificing its stored quantum information?
Bosonic quantum memories implemented in three-dimensional superconducting cavities~\cite{reagor2016,milul,paik,reagor2013} and trapped-ion motion~\cite{fluhmann} provide a natural testbed for this question.
Recent experiments have demonstrated cat-state generation~\cite{vlastakis} and dissipative confinement~\cite{leghtas2015},
quantum error correction across cat, binomial, and grid codes~\cite{ofek,hu,campagne}, beyond-break-even logical lifetimes~\cite{ni,sivak}, and fault-tolerant concatenation~\cite{putterman}.
% Sensing the field acting directly on an occupied memory would enable in-situ calibration of spatial mode profiles and background-drive drifts without transferring quantum information to auxiliary modes, tracking slowly varying fields while preserving the stored state and its reference entanglement.
Using an active bosonic memory to directly sense its local environment would enable in-situ calibration of local fields and background-drive drifts without transferring quantum information to auxiliary modes,
tracking slowly varying fields while preserving the stored state.% and its reference entanglement.

Realizing this capability requires evaluating the complete measurement instrument,
which must extract a high-contrast signal record while preserving logical recoverability.
% While quantum-enhanced metrology exploits nonclassical coherence~\cite{caves} within fundamental dissipative limits~\cite{escher,demkowicz},
Within fundamental dissipative limits~\cite{caves,escher,demkowicz},
error-corrected sensing isolates signal accumulation from specific noise channels~\cite{arrad,kessler,dur,hnls}.
Approximate quantum error correction also offers a route to optimizing precision under Markovian noise~\cite{zhou2020aqec}.
%  combining sensing with active storage imposes a stricter requirement.
% quantum memories~\cite{steane} and subsystem codes~\cite{kribs,poulin} protect logical information while accessing other degrees of freedom.
% For an endpoint readout, the oscillator operator $O$ sampled by the probe need only satisfy $POP=cP$ on the code projector $P$, in place of $O=cI$ on the whole Hilbert space.
% Continuous dissipation must preserve this readout property and recoverable storage.
%
Akin to subsystem codes~\cite{steane,kribs,poulin} that separate protected logical registers from gauge degrees of freedom,
an endpoint probe sampling an oscillator observable $O$ need only satisfy $POP=cP$ on the code projector $P$,
% in place of  $O=cI$ on the whole Hilbert space,
rather than acting trivially $O=cI$ on the entire Hilbert space.
% Under continuous dissipation, however, the open-system dynamics must simultaneously preserve this state-independent readout and maintain recoverability of the logical memory.

% Finite superpositions of equally spaced squeezed wave packets have also
% been studied as bosonic encodings~\cite{shukla2021comb}.
% Subsystem error correction has also been developed for noisy quantum
% metrology~\cite{liu2026subsystemmetrology}.

Existing bosonic architectures provide key ingredients for this separation:
grid and modular codes~\cite{gkp,terhal,royer,modular} decompose oscillator phase space into logical and gauge coordinates,
whereas binomial~\cite{binomial}, cat, and rotation-symmetric codes~\cite{cochrane,leghtas2013,mirrahimi,grimsmo,albert} protect against excitation loss.
% Finite-energy grid codes have also been optimized for correctability against photon loss and dephasing~\cite{zeng2025gkp}.
Finite-energy bosonic codes have also been designed for approximate correction of photon loss and dephasing~\cite{zeng2025gkp,zeng2026squeezedfock}.
Recent studies have explored displacement sensing and its stabilized implementations~\cite{duivenvoorden2017sensor,fluhmann2018modular,
valahu2025multiparameter,labarca}, bosonic radiometry via approximate quantum error correction~\cite{wang2022radiometry},
syndrome-based noise estimation~\cite{wagner},
entanglement-assisted phase sensing~\cite{zheng},
and state-independent geometric-phase sensing in noisy qubit-oscillator systems~\cite{suri2025}. 
A distinct question is whether force sensing can coexist with active recovery of an arbitrary stored qubit under continuous bosonic loss.
% While these approaches achieve state-independent signal extraction or robust displacement estimation,
% it remains open whether force sensing is compatible with active recovery of an arbitrary stored qubit under continuous bosonic loss.
% Here, the oscillator accumulates the signal without a sensing ancilla: we combine disjoint modular wave packets of spacing $L$ with binomial site weights.
% Within each packet, $x=LJ+y$ separates the discrete site coordinate $J$ from the force-sensitive internal coordinate $y$.
% The annihilation operator $a=(LJ+y+\partial_y)/\sqrt2$ couples these degrees of freedom, so continuous photon loss distorts both and can correlate the recovered memory with the signal probe.
% Correctability of isolated loss events alone therefore does not guarantee a compatible measurement instrument.
%

% A natural route to address this question is to combine disjoint modular wave packets of spacing $L$ with binomial site weights,
% These insights stimulate a natural route to address this question by combining disjoint modular wave packets of spacing $L$ with binomial site weights,
Addressing this question motivates us to pursue a natural route by combining disjoint modular wave packets of spacing $L$ with binomial site weights,
achieving force accumulation without a sensing ancilla during the interrogation interval.
Within each cell, decomposing position as $x = LJ + y$ separates the discrete logical site index $J$ from the force-sensitive internal coordinate $y$.
Because the annihilation operator $a = (LJ + y + \partial_y)/\sqrt{2}$ couples $J$ and $y$, continuous photon loss
simultaneously distorts the internal wave packet and induces correlations between the recovered memory and the readout probe.
Compatibility therefore requires suppressing correlations between the recovered memory and the probe at the level of the joint output channel.

% Here, we achieve ancilla-free force accumulation under continuous photon loss by combining disjoint modular wave packets of spacing $L$ with binomial site weights.
% Within each cell, the position coordinate $x = LJ + y$ separates the discrete logical site index $J$ from the force-sensitive internal coordinate $y$.
% Because the annihilation operator $a = (LJ + y + \partial_y)/\sqrt{2}$ couples $J$ and $y$, continuous photon loss simultaneously distorts the internal wave packet and induces deleterious correlations between the recovered memory and the readout probe.
% Correcting isolated photon jumps alone is therefore insufficient; a compatible sensing instrument requires joint decoupling of the memory and probe channels.  

In this Letter, we establish this joint compatibility under continuous loss $\gamma = \kappa T$.
Focusing first on a four-packet code under ideal endpoint controls,
we show that binomial site recovery cancels all packet-dependent distortions in the probe-valued loss kernel through $O(\gamma^2)$,
leaving a leading logical dephasing error of $O(\gamma^2)$ while suppressing memory-probe correlations to $O_\diamond(\gamma^3)$.
By matching higher-order site moments across $2r+2$ packets, we prove analytically that this hierarchy generalizes to arbitrary order $r$,
yielding a memory infidelity of $O(\gamma^{r+1})$ and a factorized joint-channel remainder of $O_\diamond(\gamma^{r+2})$.
% The simulations with four to ten packets verify the memory-infidelity scaling and directly resolve cubic memory-probe decorrelation for the four-packet code.
% Furthermore, we show that encodings with six or more packets eliminate first-order memory errors from intrinsic photon-number dephasing, and we evaluate ideal-control superconducting-cavity benchmarks under competing loss, dephasing, and Kerr nonlinearity.
% binomial site recovery cancels all packet-dependent distortions in the probe-valued loss kernel through $O(\gamma^2)$, leaving a leading logical dephasing error of $O(\gamma^2)$ while suppressing nonproduct memory-probe correlations to $O_\diamond(\gamma^3)$. 
% By matching higher-order site moments across $2r+2$ packets, we generalize this hierarchy to arbitrary order $r$, achieving a leading memory infidelity of $O(\gamma^{r+1})$ and a factorized joint-channel remainder of $O_\diamond(\gamma^{r+2})$. 
% Exact physical-channel simulations across four to ten packets validate this scaling and resolve cubic decorrelation for the four-packet code. 
Furthermore, we show that encodings with six or more packets eliminate first-order memory errors from intrinsic photon-number dephasing, and we benchmark the protocol against realistic superconducting-cavity noise to guide hardware implementation.

\begin{figure}[t]
 \includegraphics[width=8cm]{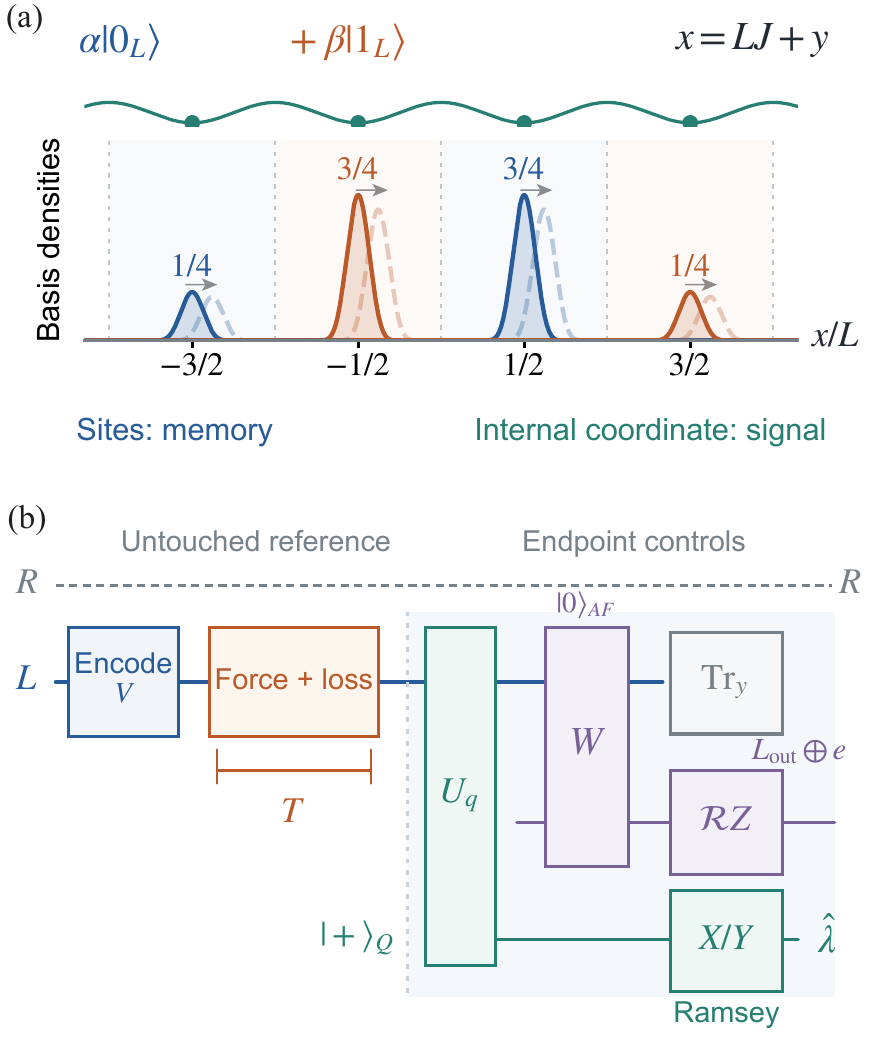}
 \caption{
   Simultaneous force sensing and recoverable quantum storage in a single bosonic mode.
   (a) Logical qubit amplitudes are encoded across discrete wave-packet sites with binomial weights ($1/4$ and $3/4$),
   while a shared internal coordinate $y$ ($x = LJ + y$) captures the force-induced displacement, illustrated by dashed lines.
   The green curve illustrates the periodic phase factor $e^{-2\pi i x/L}$,
   ensuring that the endpoint probe readout is strictly independent of the logical state.
  %  (b) Protocol circuit. An unknown logical qubit, possibly entangled with an untouched reference $R$,
  %  is encoded into the oscillator by $V$ and exposed to a force under continuous photon loss for a duration $T$.
  %  At the endpoint, a loss-calibrated conditional kick $U_q$ maps the displacement signal onto a probe qubit $Q$,
  %  initialized in $|+\rangle_Q$ and read out by Ramsey $X/Y$ measurements to estimate the force $\lambda$.
  %  Coherent site extraction $W$ transfers the cell label to an ancillary register.
  %  Tracing out the internal coordinate $y$, undoing the known logical phase $Z$, and applying polynomial site recovery $\mathcal{R}$
  %  yield the recovered memory $L_{\mathrm{out}}$, with flagged erasures $e$ retained. Endpoint controls are assumed ideal.
  (b) Protocol circuit.
  While retaining entanglement with an untouched reference $R$, the encoded oscillator state ($V$) accumulates a force signal under continuous photon loss over duration $T$.
  Under endpoint controls (shaded box), a conditional kick $U_{q_\eta}$ transfers the displacement signal to a probe qubit $\vert{}+\rangle_Q$ (measured via Ramsey $X/Y$ interferometry to obtain $\hat{\lambda}$),
   while coherent site extraction $W$ (with ancillary register $\vert{}0\rangle_{A,F}$), internal-coordinate trace (Trace $y$), and phase-corrected polynomial recovery ($Z + \mathcal{R}$) return the protected state to the memory register $L_{\text{out}} \oplus e$ (retaining flagged erasures $e$). 
 }\label{fig:scheme}
\end{figure}

\textcolor{blue}{\emph{Encoding and continuous dynamics.---}}%
The four-packet code illustrates how discrete site amplitudes store a logical qubit
while a shared internal coordinate captures the force signal, as shown in Fig. \ref{fig:scheme}(a).
Setting $[x, p] = i$, $a = (x + ip)/\sqrt{2}$, and $n = a^\dagger a$, we choose a real, even, normalized envelope $g(y)$ supported on $\vert{}y\vert{} < w/2$ with $w < L$.
The mutually orthogonal wave packets $\vert{}j, g\rangle$ have position functions $g(x - Lc_j)$ centered at $c_j = j - 3/2$ for $j \in \{0, 1, 2, 3\}$.
The logical basis states are encoded as
\begin{align}
 \ket{0_L}&=\tfrac12(\ket{0,g}+\sqrt3\ket{2,g}),\nonumber\\
 \ket{1_L}&=\tfrac12(\sqrt3\ket{1,g}+\ket{3,g}).\label{eq:code}
\end{align}
Within each cell, $x=LJ+y$ with $J=\operatorname{diag}(c_j)$.
% The two columns of site amplitudes define $V$, which satisfies
The encoding isometry $V$ defined by the site amplitudes satisfies the moment-matching conditions
\begin{equation}
 V^\dagger JV=0,\qquad V^\dagger J^2V=\tfrac34 I.\label{eq:moments}
\end{equation}
% Thus the two logical states have matched site moments, enabling recovery of site errors, while their identical internal packet responds to the force.
These matched site moments protect the logical qubit against photon loss, while the common internal wave packet $g(y)$ responds uniformly to external displacements.
The initial mean occupation is $\bar n=(3L^2/4+\langle y^2\rangle_g+\langle p^2\rangle_g-1)/2$, independent of the logical input.
% \mainrev{A compact envelope $g\propto\cos^4(\pi y/w)$ supplies finite energy and the required regularity;
% Sec.~I~A of the Supplemental Material (SM) gives the general construction~\cite{supp}.}
Using compactly supported smooth envelopes $g(y) \propto \cos^{2r+2}(\pi y/w)$ ensures both finite energy and the Sobolev regularity required for high-order error cancellation, see Supplemental Material \cite{supp}.

During the sensing interval $T$,
% the oscillator experiences a constant force $\lambda$
the oscillator experiences a constant linear displacement drive $\lambda p$
(representing an external force or microwave drive in the rotating frame)
alongside continuous photon loss at rate $\kappa$:
\begin{equation}
 \dot\rho=-i[\lambda p,\rho]+\kappa\cD[a]\rho,\quad
 \cD[a]\rho=a\rho a^\dagger-\tfrac12\{a^\dagger a,\rho\}.\label{eq:master}
\end{equation}
% Here $\lambda$ generates translation in the dimensionless coordinate $x$.
% At time $T$, the exact channel is a vacuum attenuator with $\eta=e^{-\kappa T}$ followed by a displacement $d=\lambda\tau_\kappa$, where $\tau_\kappa=2(1-\sqrt\eta)/\kappa$ and $\tau_0=T$.
% A fresh probe initialized in $\ket{+}$ then undergoes $U_q=e^{-iq_\eta\sigma_zx}$, with calibrated wave number $q_\eta=\pi/(\sqrt\eta L)$.
% The loss rate is known; encoding and endpoint controls are ideal, and $T$ counts accumulation only.
The exact time-evolved channel decomposes into a pure-loss attenuator of transmissivity $\eta = e^{-\kappa T}$ followed by a net position displacement $d = \lambda \tau_\kappa$,
with effective accumulation time $\tau_\kappa = 2(1-\sqrt{\eta})/\kappa$.
To extract the signal, a probe qubit initialized in $\vert{}+\rangle_Q$ interacts with the oscillator via the conditional kick $U_q = e^{-i q_\eta \sigma_z x}$ at the loss-calibrated wave number $q_\eta = \pi/(\sqrt{\eta}L)$.
To isolate the fundamental limits of continuous accumulation, we assume $\kappa$ is known and treat state preparation and endpoint controls as ideal impulses.

\textcolor{blue}{\emph{Exact probe readout.---}}%
Crucially, the probe extracts the force signal without acquiring any information about the logical qubit because all packet centers share an identical modular phase.
Specifically, $P e^{-2\pi i x/L} P = -c_g P$, where $P$ projects onto the code space and $c_g = \langle e^{-2\pi i y/L}\rangle_g$ is the envelope form factor.
This scalar compression yields an exact probe coherence $l_\lambda = 2\langle 0\vert{}\rho_Q\vert{}1\rangle$ after the conditional kick:
% \mainrev{The probe can respond to the force without revealing the logical input because all packet centers share the same modular phase:} $P e^{-2\pi i x/L}P=-c_gP$, where $P$ projects onto the code and $c_g=\langle e^{-2\pi i y/L}\rangle_g$.
% This scalar compression gives the probe coherence $\ell_\lambda=2\bra0\rho_Q\ket1$:
\begin{equation}
 \ell_\lambda=-c_g\exp[-(1-\eta)q_\eta^2-2iq_\eta d].\label{eq:readout}
\end{equation}
This equation holds exactly for any encoded input, including states entangled with an external reference,
across the entire $(2r+2)$-packet family, as $N = 2r+1$ is odd and all equally spaced centers share the same modular phase.
While this shared phase guarantees state-independent readout, the binomial weights ensure logical recoverability. 
The resulting probe quantum Fisher information (QFI)~\cite{braunstein} is
\begin{equation}
 F_Q=4q_\eta^2\tau_\kappa^2|c_g|^2e^{-2(1-\eta)q_\eta^2}.\label{eq:qfi}
\end{equation}
Locally optimized Ramsey detection attains this QFI,
while equally sampling two orthogonal probe quadratures guarantees a classical Fisher information of at least $F_Q/2$ per shot within the unambiguous phase interval \cite{supp}.
However, an input-independent probe marginal does not by itself guarantee recoverable storage,
as backaction from the readout kick and continuous loss can still disturb the memory or correlate it with the probe.

% Locally optimized Ramsey detection attains this QFI; equal allocation to two orthogonal quadratures guarantees at least $F_Q/2$ per shot.
% \mainrev{The prior must resolve phase wrapping; SM Sec.~I~B gives the phase range and the equal-quadrature bound~\cite{supp}.
% This exact probe marginal does not by itself establish recoverable storage: the memory may still be disturbed or correlated with the probe.}

\textcolor{blue}{\emph{Four-packet cancellation mechanism.---}}%
To evaluate storage and sensing as a unified instrument, we analyze the joint output channel $\mathcal{J}_{\lambda,\gamma}$ mapping the encoded memory to the recovered logical register and the unmeasured probe.
As depicted in Fig. \ref{fig:scheme}(b), the endpoint decoder $W$ coherently transfers the contracted cell index (of width $\sqrt{\eta}L$) into a two-qubit ancillary register,
flags any population outside the cells as an erasure qubit $e$,
and traces out the internal coordinate $y$ (see Supplemental Material \cite{supp}).
After undoing the deterministic logical phase $Z$ induced by the readout kick, the site recovery is
% \mainrev{To assess storage and sensing together, we retain the recovered memory and the unmeasured probe. The endpoint decoder coherently transfers the cell label to a register,
% then traces out the oscillator's internal coordinate [Fig.~\ref{fig:scheme}(b)].
% It uses cells of width $\sqrt\eta L$ and flags the outside region as erasure.
% All outcomes, including flagged erasures, are retained. For four packets, two label qubits and an additional flag suffice;
% a register unitary separates the logical qubit from its syndrome. SM Sec.~I~C specifies this ideal dilation and bounds cell crossings and erasures;
% smooth filters and control sequences remain to be compiled~\cite{eickbusch,terhal,supp}.}
% After undoing the kick's known logical $Z$, the site recovery is
\begin{equation}
 R_0=V^\dagger,\qquad R_1=\frac{2}{\sqrt3}V^\dagger J,\label{eq:recovery}
\end{equation}
which satisfy $\sum_bR_b^\dagger R_b=I_4$ by Eq.~\eqref{eq:moments}.
The same moment conditions make $\{I,a,a^\dagger\}$ satisfy the Knill--Laflamme criterion~\cite{kl}. % but note that finite-time analysis must also retain internal-packet distortion.
% Optimized approximate recoveries are a further possibility~\cite{barnum,beny}.

% Let $\cJ_{\lambda,\gamma}$ output the recovered logical register, including erasure, and the unmeasured probe.
% At fixed positive decoding guard $a_*=(L-w)/2-|d|/\sqrt\eta>0$, boundary corrections obey $\epsilon_{\rm bd}=O(e^{-b/\gamma})$ for some $b>0$.
% For sufficiently smooth packets and fixed $L$, we establish
Provided the displacement remains within the positive decoding guard $a_* = (L-w)/2 - \vert{}d\vert{}/\sqrt{\eta} > 0$,
cell-crossing boundary errors are exponentially suppressed as $\epsilon_{\mathrm{bd}} = O(e^{-b/\gamma})$ ($b > 0$).
For sufficiently smooth packets, the joint memory-probe channel factorizes as
\begin{equation}
 \cJ_{\lambda,\gamma}(\rho_L)=
 \big[(1-p)\rho_L+pZ\rho_LZ\big]\otimes\rho_Q
 +O_\diamond(\gamma^3)+O_\diamond(\epsilon_{\rm bd}),
 \label{eq:joint}
\end{equation}
where $p=9L^4\gamma^2/64$ and $\rho_Q$ is the exact probe state of Eq.~\eqref{eq:readout}.
% Thus the memory retains a quadratic logical dephasing error, while its correlations with the probe vanish through that order.
% \mainrev{The diamond-norm remainder applies to arbitrary reference-entangled inputs and any subsequent probe measurement.}
Thus, the memory experiences only a quadratic logical dephasing error $O(\gamma^2)$, while deleterious memory-probe correlations are suppressed to $O_\diamond(\gamma^3)$ for arbitrary reference-entangled inputs and subsequent probe measurements.
The derivation is detailed in the End Matter.

% After removal of the known site $Z$, the mechanism is visible in the full probe-valued kernel for site separation $m$:
The physical origin of this third-order decoupling is revealed by expanding the probe-valued loss kernel $Q_m$ for site separation $m = j - k$:
\begin{equation}
 Q_m=\rho_Q+m^2B_2(\gamma)+\frac{L^4\gamma^2}{32}m^4\rho_Q+O(\gamma^3).
 \label{eq:physicalcancel}
\end{equation}
Here $B_2(\gamma)=O(\gamma)$ encapsulates all packet-dependent distortions through second order.
Crucially, the recovery operation combines these off-diagonal kernels as $(9Q_1-Q_3)/8$.
Because $9\cdot 1^2 - 3^2 = 0$, this linear combination identically eliminates the entire $m^2 B_2(\gamma)$ term,
canceling all internal wave-packet distortions through second order,
while the surviving $m^4$ contribution sets the leading scalar dephasing $p$.
Consequently, the envelope shape $g(y)$ governs the sensing contrast,
whereas the leading memory infidelity depends solely on the lattice spacing $L$.
Tracing out the probe yields a reduced entanglement infidelity~\cite{schumacher} of $1 - F_e = p + O(\gamma^3) + O(\epsilon_{\mathrm{bd}})$,
representing a quadratic improvement over unrecovered code-space projection, which scales linearly as $3L^2\gamma/8 + O(\gamma^2)$~\cite{supp}.   

% Since $9\cdot1^2-3^2=0$, it cancels every $m^2B_2$ term, including internal-packet distortion through second order, while the $\gamma^2m^4\rho_Q$ contribution fixes the leading dephasing.
% Packet shape therefore determines signal contrast, whereas the leading memory coefficient depends only on $L$ within the stated assumptions.
% Tracing the probe gives the entanglement infidelity~\cite{schumacher} $1-F_e=p+O(\gamma^3)+O(\epsilon_{\rm bd})$; projection alone, with discarded events counted as erasures, instead gives $3L^2\gamma/8+O(\gamma^2)$.
% \mainrev{SM Sec.~II~A derives the scalar fidelity and probe-valued cancellation formulas~\cite{supp}.}

\textcolor{blue}{\emph{Generalization to higher correction order.---}}%
This cancellation mechanism extends systematically to arbitrary order $r \ge 1$ by matching higher site moments across $N+1 = 2r+2$ wave packets centered at $c_j = j - N/2$ ($j = 0, \dots, N$, with $N = 2r+1$):
% For $N=2r+1$, $r\geq1$, use $N+1=2r+2$ packets with $c_j=j-N/2$, $j=0,\ldots,N$, and logical labels $\mu=0,1$:
\begin{equation}
 \ket{\mu_L}=2^{-(N-1)/2}
 \sum_{j\equiv\mu\ ({\rm mod}\ 2)}\sqrt{\binom Nj}\ket{j,g}.
 \label{eq:hierarchy}
\end{equation}
% Binomial cancellation makes $V^\dagger J^kV$ scalar for $k\leq2r$.
% Orthogonalizing $V,JV,\ldots,J^rV$ gives a complete site recovery.
% \mainrev{As derived in SM Sec.~II~B, a random-rotation representation maps the reduced-memory infidelity to a repetition-code binomial tail~\cite{supp}, giving}
Binomial identities ensure that $V^\dagger J^k V$ is proportional to the identity for all $k \le 2r$, allowing Gram-Schmidt orthogonalization of $\{V, JV, \dots, J^r V\}$ to define a complete site recovery.
Mapping the reduced-memory channel onto a classical random-rotation mixture yields a binomial-tail entanglement infidelity \cite{supp}:
\begin{equation}
 1-F_e=\frac{\binom{2r+1}{r+1}(2r+1)!!}{8^{r+1}}
 L^{2r+2}\gamma^{r+1}+O(\gamma^{r+2})+O(\epsilon_{\rm bd}).\label{eq:order}
\end{equation}
Retaining the full probe-valued kernels yields the recovered off-diagonal operator $H_r = \sum_{k=0}^r c_{r,k} Q_{2k+1}$, where the Lagrange interpolation weights $c_{r,k}$ at nodes $(2k+1)^2$ satisfy $\sum_{k=0}^r c_{r,k}(2k+1)^{2a} = 0$ for all $1 \le a \le r$. These algebraic identities cancel all packet-dependent probe distortions through order $\gamma^{r+1}$~\cite{supp}.
Denoting the leading dephasing probability in Eq.~\eqref{eq:order} by $\pi_r$ and defining $\mathcal{D}_\pi(\rho) = (1-\pi)\rho + \pi Z\rho Z$, we obtain the general factorization theorem:
\begin{equation}
 \cJ_r(\rho)=\cD_{\pi_r}(\rho)\otimes\rho_Q
 +O_\diamond(\gamma^{r+2})+O_\diamond(\epsilon_{\rm bd}),
 \label{eq:generaljoint}
\end{equation}
which holds at any fixed order $r$, spacing $L$, readout harmonic $h$, and positive guard $a_* > 0$ for compactly supported even envelopes $g \in H^{r+2}(\mathbb{R})$.

\begin{figure}[t]
\includegraphics[width=\columnwidth]{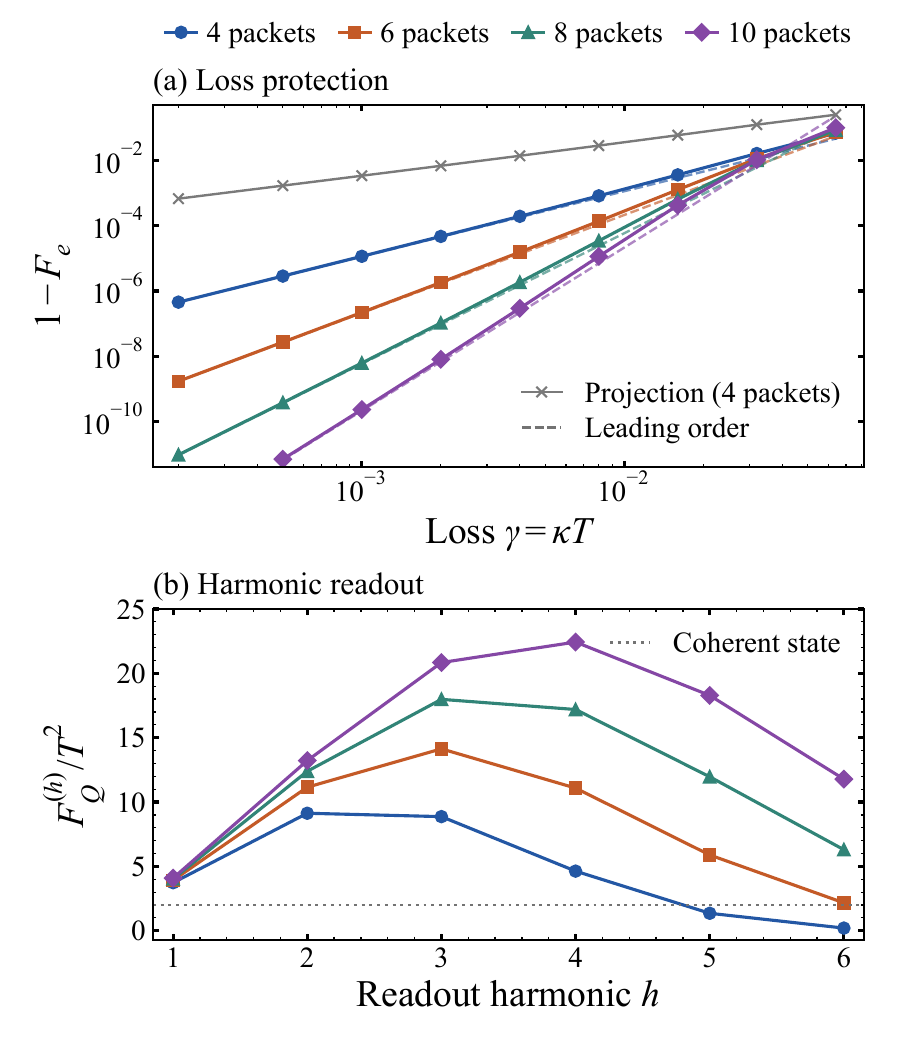}
 \caption{
   Performance under continuous photon loss for four through ten wave packets
   using envelopes $g(y) \propto \cos^{2r+2}(\pi y/w)$ at $L = 3$ and $w = 1.8$.
   (a) Exact physical-channel memory entanglement infidelity $1 - F_e$ versus loss $\gamma = \kappa T$ at $\lambda T = 0.2$ and $h = 1$.
   Dashed lines show the analytical leading-order predictions of Eq. (10);
   gray crosses indicate unrecovered four-packet code projection (including erasures).
   (b) Normalized probe QFI $F_Q^{(h)}/T^2$ versus readout harmonic $h$ at $\gamma = 10^{-3}$,
   compared against the coherent-state benchmark $2\tau_\kappa^2/T^2 \simeq 2$ (dotted line).
   Tighter spatial localization of higher-order envelopes slows form-factor decay, shifting the optimal readout to higher harmonics.
   % Loss-model performance at $L=3$, $w=1.8$ for $g(y)\propto\cos^{2r+2}(\pi y/w)$ and four through ten packets. (a) Exact physical-channel memory infidelity at $\lambda T=0.2$, $h=1$. Gray crosses: four-packet projection with erasures; dashed lines: leading terms in Eq.~\eqref{eq:order}. Joint tests appear in the End Matter. (b) Exact probe QFI at $\gamma=10^{-3}$. The gray dotted line is the coherent-state displacement reference $2\tau_\kappa^2/T^2\simeq2$. The narrower higher-power envelopes favor higher harmonics; packet shapes and energies vary across curves. The common $h=2$ readout is unambiguous throughout the protected interval.
   }
    \label{fig:results}
\end{figure}

To validate the analytical scaling in Eq.~\eqref{eq:order}, as shown in Fig.~\ref{fig:results}(a), we numerically simulate the memory infidelity $1-F_e$ for codes with four through ten packets using envelopes
$g(y) \propto \cos^{2r+2}(\pi y/w)$ at $L = 3$ and $w = 1.8$, 
corresponding to initial photon numbers $\bar{n} = 6.37, 10.12, 13.88$, and $17.65$,
clearly resolving quadratic through quintic suppression of memory infidelity.
At $\gamma = 10^{-3}$ and $\lambda T = 0.2$, the four-packet code already achieves $1 - F_e = 1.16 \times 10^{-5}$.
As shown in Fig.~\ref{fig:jointend} in the End Matter, we evaluate the complete reference-entangled joint channel,
confirming the same leading scaling orders and directly resolving the $O(\gamma^3)$ memory-probe decorrelation for four packets.
Moreover, the same binomial moment conditions suppress logical state contamination from relative readout-kick calibration errors $\epsilon$ to $O(\vert{}\epsilon\vert{}^{2r+1})$ \cite{supp}.

% \mainrev{Figure~\ref{fig:results}(a) tests Eq.~\eqref{eq:order} for four through ten packets,
% resolving quadratic through quintic reduced-memory suppression.
% The End Matter tests the complete memory--probe channel with a reference, erasures, and cell crossings retained [Fig.~\ref{fig:jointend}]. Its joint distances support the same leading orders in the resolved loss windows; the stronger cubic remainder is tested separately for four packets.}

% For this numerical family, $g(y)\propto\cos^{2r+2}(\pi y/w)$ inside its support provides the required regularity.
% Both packet count and shape change, with initial energies $\bar n=6.37,10.12,13.88,17.65$ at $L=3$, $w=1.8$.
% Increasing $r$ also increases the recovery-register dimension.
% \mainrev{At $\gamma=10^{-3}$ and $\lambda T=0.2$, four packets give $1-F_e=1.16\times10^{-5}$. SM Sec.~V and Table~S2 collect the hierarchy data and independent Fock-basis readout checks~\cite{supp}. The same moment construction suppresses logical contamination from a relative kick error $\varepsilon$ as $O(|\varepsilon|^{2r+1})$; Sec.~II~C derives this bound, which does not imply the same scaling for force-estimation bias~\cite{supp}.}

\textcolor{blue}{\emph{Harmonic choice and sensing range.---}}%
The shared modular phase also allows us to boost force sensitivity by reading out at a higher integer harmonic $h > 1$ without compromising loss protection. Applying the kick at wave number $q_{\eta,h} = h\pi/(\sqrt{\eta}L)$
and undoing the deterministic logical phase $Z^h$ prior to recovery preserves exact state-independent readout, yielding a harmonic QFI of
% \mainrev{The shared modular phase also permits optimization of the readout harmonic while preserving the fixed-order loss protection.}
% For an integer $h\geq1$, set $q_{\eta,h}=h\pi/(\sqrt\eta L)$ and $c_{g,h}=\langle e^{-2\pi i h y/L}\rangle_g$, undoing the known site operation $Z^h$ before recovery.
% The probe remains exactly input independent, with
\begin{equation}
 F_Q^{(h)}=4q_{\eta,h}^2\tau_\kappa^2|c_{g,h}|^2
 e^{-2(1-\eta)q_{\eta,h}^2},
 \label{eq:harmonicqfi}
\end{equation}
where $c_{g,h} = \langle e^{-2\pi i h y/L}\rangle_g$.
At every fixed $h$, Eq.~\eqref{eq:generaljoint} preserves the leading memory error throughout the family; the reduced-memory channel is exactly independent of $h$ in the original-label comparison model~\cite{supp}.

As shown in Fig.~\ref{fig:results}(b),
% the phase sensitivity illustrates the trade-off between the enhanced phase slope $\propto h^2$ and the decay of both the envelope form factor $\vert{}c_{g,h}\vert{}$ and loss visibility.
the linear growth of the phase slope with $h$ contributes an $h^2$ factor to $F_Q^{(h)}$ that competes against form-factor and loss-visibility decay.
Because the higher-order envelopes $\cos^{2r+2}(\pi y/w)$ are more tightly localized in position, $\vert{}c_{g,h}\vert{}$ decays more slowly with $h$,
shifting the optimal readout harmonic to larger values as the code size increases.
For the four-packet code at $\gamma = 10^{-3}$, switching from $h = 1$ to $h = 2$ more than doubles the normalized sensitivity $F_Q/T^2$ from $3.73$ to $9.13$,
substantially exceeding the coherent-state displacement reference $2\tau_\kappa^2/T^2 \simeq 2$.
Because the phase period in $\lambda$ is $\sqrt{\eta}L/(h\tau_\kappa)$, choosing $h = 2$ at $L = 3$ and $w = 1.8$ satisfies $2(L-w) < L$, ensuring unambiguous phase estimation across the entire protected guard interval.
% We therefore adopt $h = 2$ for our device benchmarks below.

% Figure~\ref{fig:results}(b) shows the competition between increasing phase slope and decreasing packet contrast and loss visibility.
% The chosen higher-power envelopes are more localized, slowing the decay of $|c_{g,h}|$ and favoring higher optimal harmonics.
% This trend reflects the specified change in envelope along with code size, rather than a sensitivity gain due to correction order alone.

% For the same four-packet encoding at $\gamma=10^{-3}$, changing $h=1$ to $2$ raises $F_Q/T^2$ from $3.73$ to $9.13$.
% The coherent-state displacement scale $2\tau_\kappa^2/T^2\simeq2$ in Fig.~\ref{fig:results}(b) is a reference, not a comparison at matched sensor--memory resources.
% The phase period in $\lambda$ is $\sqrt\eta L/(h\tau_\kappa)$; $h=2$ covers the full protected interval for $L=3$, $w=1.8$ because $2(L-w)<L$.
% We use this common harmonic below.
% Higher $h$ can improve local QFI but may require a narrower prior or coarse estimation, and increases the endpoint kick amplitude and its control cost.
% The fixed-$h$ expansion does not cover harmonics increasing with $1/\gamma$.

% \textcolor{blue}{\emph{Intrinsic number dephasing.---}}%
Notably, beyond photon loss, practical bosonic memories experience intrinsic photon-number dephasing generated by $(2/T_\phi)\mathcal{D}[n]$.
In the four-packet code, the recovery operator $R_1$ maps part of the $J^2$ component of $n$ into a logical $Z$ error, leaving a linear memory infidelity:
% The vacuum-loss theorem provides a baseline, but additional noise need not respect the same hierarchy. For intrinsic number dephasing generated by $2\mathcal D[n]/T_\phi$, four packets retain the memory error
\begin{equation}
 (1-F_e)_\phi=\frac{3L^4}{8}\frac{T}{T_\phi}+O((T/T_\phi)^2),\label{eq:dephasing}
\end{equation}
Crucially, increasing the code size to six or more packets matches site moments through degree four,
% placing $I$, $J$, and $J^2$ entirely within orthogonal, correctable syndrome subspaces.
resolving the error space generated by $I$, $J$, and $J^2$ into mutually orthogonal syndrome subspaces.
This eliminates the first-order dephasing error, suppressing pure number-dephasing infidelity to $O((T/T_\phi)^2)$ (with leading mixed loss-dephasing corrections entering at $O(\kappa T^2/T_\phi)$~\cite{supp}).
Exact Fock-space simulations under pure dephasing confirm this transition, yielding fitted scaling exponents of $1.000$ and $1.997$ for four and six packets, respectively.

% because the recovery maps part of the $J^2$ component of $n$ into logical $Z$. Six or more packets match moments through degree four, placing $I,J,J^2$ in correctable syndrome spaces and removing the first-order memory error. Pure number dephasing then gives $O((T/T_\phi)^2)$; independent four- and six-packet fits yield exponents $1.000$ and $1.997$. Mixed loss--dephasing terms can remain at order $(\kappa T)(T/T_\phi)$. SM Sec.~III proves this first-order correction under its stated regularity assumptions; it does not extend the arbitrary-order joint-channel theorem to number dephasing~\cite{supp}.

\begin{figure}[t]
 \includegraphics[width=\columnwidth]{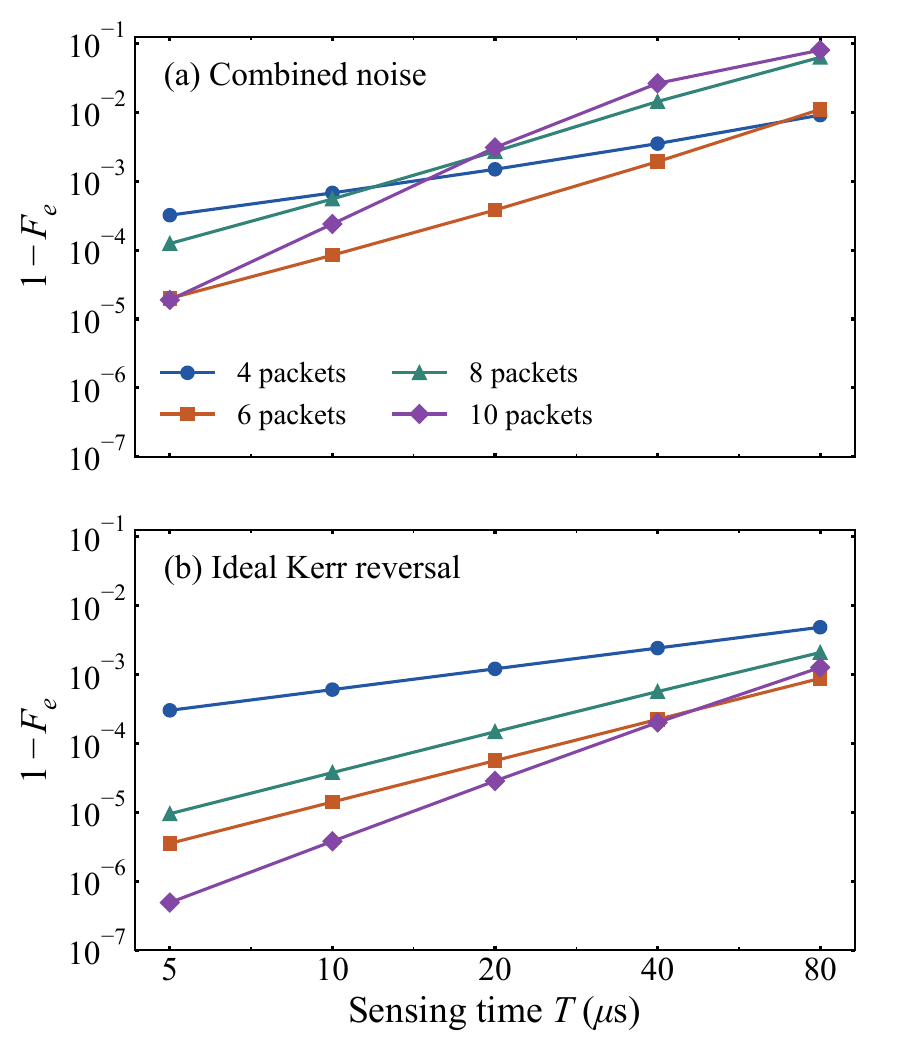}
 \caption{
   Cavity memory benchmarks showing entanglement infidelity $1 - F_e$ versus sensing duration $T$
   under (a) combined photon loss ($T_1^c = 25.6~\mathrm{ms}$), residual Kerr nonlinearity ($K/2\pi = 3.6~\mathrm{Hz}$),
   and intrinsic number dephasing ($T_\phi = 0.5~\mathrm{s}$),
   and (b) the same evolution followed by endpoint Kerr reversal, evaluated at $\lambda T = 0.2$ and $h = 2$.
   At $T = 20~\mu\mathrm{s}$, the six-packet code is optimal without compensation, whereas the ten-packet code performs best with Kerr reversal.
   % Memory entanglement infidelity under (a) loss, Kerr, and intrinsic number dephasing and (b) the same noise with ideal endpoint Kerr reversal. Parameters: $T_1^c=25.6$ ms, $K/2\pi=3.6$ Hz, stress-test $T_\phi=0.5$ s, $\lambda T=0.2$, $h=2$. Colors and markers identify four through ten packets; lines guide the eye. At $20\,\mu$s, six packets perform best in (a), ten in (b). Endpoint controls are ideal; thermal transmon events and control faults are excluded.
   }\label{fig:practical}
\end{figure}

\textcolor{blue}{\emph{Competing noise in cavity benchmarks.---}}%
While larger packet counts suppress photon loss and number dephasing, their higher mean photon numbers and number variances increase susceptibility to residual Kerr nonlinearity.
To determine the optimal encoding under realistic hardware constraints, we benchmark
our hierarchy using parameters from a state-of-the-art dispersive cavity-transmon system~\cite{milul}.
Such circuit-QED systems~\cite{blais2004,wallraff,koch,blais2021,you2011circuit,gu2017microwave} support fast echoed conditional displacements (ECDs) and number-selective phase gates~\cite{heeres2015,krastanov,heeres2017,eickbusch}.
% To determine the optimal encoding under realistic hardware constraints, we benchmark
% our hierarchy using parameters from a state-of-the-art dispersive cavity-transmon system~\cite{milul,blais2004,wallraff,koch,blais2021},
% which supports fast echoed conditional displacements (ECDs) and number-selective phase gates~\cite{heeres2015,krastanov,heeres2017,eickbusch}.
%
% \mainrev{We now ask which encoding remains favorable when intrinsic number dephasing competes with residual Kerr evolution. The increasing energies and number moments of this family also increase its sensitivity to Kerr.}
% A dispersive cavity--transmon system~\cite{blais2004,wallraff,koch,blais2021} provides control tools including displacements, number-selective phases, and echoed conditional displacements~\cite{heeres2015,krastanov,heeres2017,eickbusch}.
% \mainrev{Using the reported cavity lifetime $T_1^c=25.6$ ms and simulation-derived $K/2\pi=3.6$ Hz for one device~\cite{milul}, we stress-test intrinsic dephasing by setting $T_\phi=0.5$ s at the reported lower bound:}
Adopting the reported cavity lifetime $T_1^c = 25.6~\mathrm{ms}$, residual Kerr coefficient $K/2\pi = 3.6~\mathrm{Hz}$, and a conservative lower-bound dephasing time $T_\phi = 0.5~\mathrm{s}$~\cite{milul}, we simulate the full master equation:
\begin{equation}
 \dot\rho=-i[\lambda p-\tfrac K2n(n-1),\rho]
 +\frac{\cD[a]\rho}{T_1^c}+\frac{2\cD[n]\rho}{T_\phi}.\label{eq:practical}
\end{equation}
% \mainrev{Figure~\ref{fig:practical} compares reduced-memory entanglement infidelity for four through ten packets at the common harmonic $h=2$: (a) under combined noise and (b) after ideal, signal-independent endpoint Kerr reversal. Both retain an entangled reference and use converged Fock calculations with ideal endpoint controls. SM Sec.~IV specifies the device model; Sec.~V gives convergence checks and the loss-only position-space reference~\cite{supp}.}
The corresponding simulation results are shown in Fig. \ref{fig:practical}, with the reduced-memory entanglement infidelity at $h = 2$ and $\lambda T = 0.2$ for four through ten packets.
At $T = 20~\mu\mathrm{s}$, the six-packet code gives the lowest infidelity among the tested encodings without Kerr compensation, as shown in Fig.~\ref{fig:practical}(a), balancing first-order dephasing suppression against lower Kerr sensitivity.
Endpoint Kerr reversal makes the ten-packet code preferable, with $1-F_e=2.86\times10^{-5}$ as shown in Fig.~\ref{fig:practical}(b).
Notably, even after Kerr reversal, the six-packet code outperforms the eight-packet code and overtakes the ten-packet code at longer durations ($T = 80~\mu\mathrm{s}$).
% Controlled ablation studies with $K = 0$ and pure number dephasing \cite{supp} confirm that this non-monotonic ordering arises directly from higher-order number-dephasing moments rather than residual Kerr cross-terms. 
Controlled ablation studies setting $K = 0$ and isolating pure $\mathcal{D}[n]$ \cite{supp} show that higher-order number-dephasing moments alone are sufficient to reproduce this non-monotonic ordering.

Under Kerr and number dephasing, the probe readout acquires a weak dependence on the logical input.
Although the calculated linearized local calibration bias remains below $3.1 \times 10^{-5}$ at $T = 20\ \mu\text{s}$ for the tested codes,
treating an unknown logical Bloch coordinate as a nuisance parameter across repeated state preparations can substantially reduce the profiled Fisher information due to near-collinearity between the signal and nuisance tangents
(whereas multi-cycle sensing with a single preserved state avoids repeated unknown-state preparation; see Supplemental Material \cite{supp}).
Finally, practical hardware implementation must account for endpoint control overheads and ancillary transmon faults:
while an ECD readout kick requires only $1.1\text{--}2.2\ \mu\text{s}$,
a parasitic transmon thermal jump during $T = 20\ \mu\text{s}$ occurs with probability $2.17 \times 10^{-4}$ for $T_\uparrow^q = 92\text{ ms}$~\cite{milul}, exceeding the ideal Kerr-reversed ten-packet memory error,
and the spatial decoder $W$ requires compiling finite-bandwidth extraction pulses.
% Residual Kerr nonlinearity also weakly couples the probe readout to the logical state.
% We evaluate both the worst-case conditional Fisher information across fixed Pauli inputs and the profiled Fisher information treating a single logical coordinate as an unknown nuisance parameter,
% showing that finite-shot estimation bias remains below $3.1 \times 10^{-5}$ at $T = 20~\mu\mathrm{s}$~\cite{supp}.
% Regarding endpoint control overheads, implementing the $h = 2$ readout kick via ECD gates with intermediate cavity displacements $\vert{}\alpha_0\vert{} \in [5, 10]$ requires an interaction time of only $1.1\text{--}2.2~\mu\mathrm{s}$.
% Furthermore, with the ancillary transmon initialized in its ground state, its reported heating time of $T_\uparrow^q = 92~\mathrm{ms}$~\cite{milul} bounds the probability of a parasitic thermal jump during a $20~\mu\mathrm{s}$ sensing window to $2.17 \times 10^{-4}$.
Incorporating full pulse compilation and fault-tolerant syndrome extraction~\cite{rosenblum} represents a natural next step toward experimental realization.  

% \mainrev{Kerr evolution also weakly couples probe readout to the logical input. SM Sec.~IV~A and Fig.~S1 distinguish the minimum conditional Fisher information over six fixed Pauli inputs from information profiled over one unknown logical component of identically prepared inputs~\cite{supp}. Even weak contamination can cause a large profiling penalty if its coherence tangent aligns with the signal tangent; finite-shot bias is assessed separately. These diagnostics do not characterize an arbitrary unknown logical state.

% Control costs require a separate budget. The second-harmonic kick has an estimated ideal echoed-conditional-displacement interaction time of $1.1$--$2.2\,\mu$s at intermediate amplitudes $5$--$10$. With an initial ground-state reset, the reported transmon heating time of $92$ ms gives event probability $2.17\times10^{-4}$ at $20\,\mu$s~\cite{milul}. SM Sec.~IV~B converts this probability into an additional error bound only under state-independent Poisson excitation and a baseline no-event branch~\cite{supp}. Thermal events, pulse and decoder overheads, and auxiliary faults~\cite{rosenblum} are excluded from the plotted channels and must enter a full implementation budget.}

\textcolor{blue}{\emph{Conclusions.---}}%
% \mainrev{Finite-energy wave packets place the stored qubit in site amplitudes and the force signal in a common internal coordinate. A shared modular phase gives exact input-independent readout under vacuum loss; probe-valued moment cancellation suppresses memory--probe correlations one order beyond the leading logical dephasing at every fixed correction order. Physical-channel calculations test the first four members of the hierarchy and resolve cubic decorrelation for four packets. Cavity benchmarks show how number dephasing, Kerr evolution, and compensation change the preferred encoding. For repeated sensing of the same stored state, SM Sec.~IV~C analyzes recovery and re-encoding with fresh packets, memory and information budgets in the product approximation, and accumulated channel errors~\cite{supp}.}
% Placing the signal in the internal coordinate is consistent with the Lindblad-span constraint on protected logical metrology~\cite{hnls}.
% Compiling smooth register extraction, accounting for total control resources, and extending the construction to gates and concatenation~\cite{puri,noh,nohchamberland} remain steps toward sensing within an occupied quantum memory.
%
We have established a family of finite-energy bosonic codes that decouple quantum memory protection from continuous force sensing within a single oscillator.
By storing logical information in binomial site amplitudes while accumulating the force signal in a shared internal coordinate,
our protocol achieves exact state-independent readout under photon loss
and suppresses memory-probe correlations one order beyond the leading logical dephasing at every correction order.
% Because the displacement signal is retained in the internal coordinate rather than accumulated as a protected logical rotation, this division of labor is fully consistent with Lindblad-span constraints on error-corrected logical metrology~\cite{hnls}.
% This division of labor naturally circumvents no-go constraints on error-corrected logical metrology~\cite{hnls}
% by routing the signal through the internal gauge subsystem rather than the protected logical manifold.
% Furthermore, our framework supports multi-cycle sensing of a persistent force while retaining the same stored logical state via periodic recovery and fresh wave-packet re-encoding~\cite{supp}.
% Compiling smooth finite-bandwidth extraction pulses, mitigating logical nuisance coupling under Kerr nonlinearity, and integrating these wave-packet encodings into fault-tolerant bosonic architectures~\cite{puri,noh,nohchamberland}
% Covariant approximate codes can protect metrologically useful probe states when the signal couples to a logical rotation~\cite{lin2025covariant}.
Because the signal is retained in the internal coordinate rather than accumulated as a protected logical rotation, the protocol is consistent with Lindblad-span constraints on error-corrected logical metrology~\cite{hnls,demkowicz2017adaptive}.
More broadly, our results motivate exploring weak-field sensing with occupied bosonic memories, including potential applications to cavity searches for dark-photon dark matter, whose coupling induces a weak displacement of the microwave mode~\cite{labarca,dixit,agrawal,kang}.
Integrating these wave-packet encodings into fault-tolerant bosonic architectures~\cite{puri,noh,nohchamberland}
% will open new avenues for in-situ field calibration and quantum-enhanced metrology inside active quantum processors.
represent key next steps toward in-situ field calibration in quantum processors.

\emph{Acknowledgments.---}%
This work was partially supported by the Quantum Science and Technology--National Science and Technology Major Project (Grant Nos. 2021ZD0301702 and 2024ZD0302000),
the Natural Science Foundation of Jiangsu Province (Grant Nos. BK20232002 and BK20233001),
the Basic Research Program of Jiangsu (Grant No. BK20251248),
the National Natural Science Foundation of China (Grant Nos. 12074179 and 12504579),
the Postdoctoral Fellowship Program of CPSF (Grant No. GZC20261876),
and the Natural Science Foundation of Shandong Province (Grant No. ZR2023LZH002).

% ~\\

\emph{Data and code availability.---}%
The data that support the findings of this Letter are not publicly available. The data are available from the authors upon reasonable request.

\let\originalbibitem\bibitem
\renewcommand{\bibitem}[2][]{\ifstrequal{#2}{suri2025}{}{}\originalbibitem[#1]{#2}}
% \bibliography{references}

%apsrev4-2.bst 2019-01-14 (MD) hand-edited version of apsrev4-1.bst
%Control: key (0)
%Control: author (8) initials jnrlst
%Control: editor formatted (1) identically to author
%Control: production of article title (0) allowed
%Control: page (0) single
%Control: year (1) truncated
%Control: production of eprint (0) enabled
%

\clearpage
\raggedbottom
% REVTeX also uses its own column-bottom switch.
\makeatletter\@booleantrue\raggedcolumn@sw\makeatother
\section*{End Matter}
\appendix*
% \section{Joint-channel evidence}
\label{app:joint}
\setcounter{equation}{0}
\renewcommand{\theequation}{A\arabic{equation}}
\renewcommand{\theHequation}{A.\arabic{equation}}

\textcolor{blue}{\emph{Derivation of joint factorization.---}}%
Equation~\eqref{eq:joint} establishes that the joint memory--probe channel factorizes into a logical dephasing channel and the unperturbed probe marginal $\rho_Q$, with nonproduct memory--probe correlations suppressed to $O_\diamond(\gamma^3)$~\cite{supp}. The derivation proceeds in four steps.

Step 1: Probe-valued site kernel.
The decoder partitions the position axis into cells of width $\sqrt{\eta}L$, parameterized locally as $x = \sqrt{\eta}(Lc_j + y)$. Up to exponentially small boundary leakage $\epsilon_{\rm bd} = O(e^{-b/\gamma})$, coherence between sites $j$ and $k$ couples to the probe qubit via a probe-valued kernel $Q_m$ that depends solely on the site separation $m = j - k$:
\begin{equation}
 Q_m = \frac{e^{-u m^2}}{2}
 \begin{pmatrix}
  A_g(s_m) & \ell_m \\
  \ell_m^* & A_g(s_m)
 \end{pmatrix},
 \label{eq:emQm}
\end{equation}
where $\ell_m = (-1)^N e^{-B-i\phi} C_g(s_m)$, $u = \eta(1-\eta)L^2/4$ governs inter-site Gaussian damping, $s_m = (1-\eta)Lm$ is the loss-induced relative packet drift, $B = (1-\eta)q_\eta^2$, and $\phi = 2q_\eta d$. Here, $A_g(s) = \int dy\, g(y-s/2)g(y+s/2)$ and $C_g(s) = \int dy\, g(y-s/2)g(y+s/2)e^{-2\pi i y/L}$ denote the envelope overlap integrals. At $m = 0$, $s_0 = 0$ yields $A_g(0) = 1$ and $C_g(0) = c_g$, so the diagonal kernel reduces identically to the exact probe marginal, $Q_0 \equiv \rho_Q$.

Step 2: Small-loss expansion ($\gamma \ll 1$).
Expanding $1-\eta = \gamma - \gamma^2/2 + O(\gamma^3)$ for a smooth, even envelope $g(y)$ yields $A_g(s_m) = 1 - \langle p^2\rangle_g L^2 \gamma^2 m^2/2 + O(\gamma^3)$ and $e^{-u m^2} = 1 - L^2\gamma m^2/4 + 3L^2\gamma^2 m^2/8 + L^4\gamma^2 m^4/32 + O(\gamma^3)$. Collecting powers of $m$ gives Eq. \eqref{eq:physicalcancel}.
% \begin{equation}
%  Q_m = Q_0 + m^2 B_2(\gamma) + \frac{L^4\gamma^2}{32} m^4 Q_0 + O(\gamma^3).
%  \label{eq:emQexpand}
% \end{equation}
% where the matrix $B_2(\gamma) = O(\gamma)$ encapsulates all envelope-dependent internal distortions through second order, whereas the $m^4$ contribution remains strictly proportional to the unperturbed probe state $Q_0$.

Step 3: Cancellation under four-packet recovery.
According to Eq. \eqref{eq:recovery}, because the logical basis states have disjoint parity support, completeness preserves the diagonal logical blocks as $\rho_{00}\ket{0}\bra{0}\otimes Q_0 + \rho_{11}\ket{1}\bra{1}\otimes Q_0$. Conversely, the off-diagonal blocks couple opposite-parity sites $j \in \{0,2\}$ and $k \in \{1,3\}$. Across the four inter-site pairs, recovery assigns weights of $+3/8$ to each $|m|=1$ pair (summing to $9/8$) and $-1/8$ to the $|m|=3$ pair, yielding the effective off-diagonal kernel
\begin{align}
 H &= \frac{9Q_1 - Q_3}{8} \nonumber\\
   &= Q_0 + \frac{9\cdot 1^2 - 3^2}{8} B_2(\gamma) + \frac{9\cdot 1^4 - 3^4}{8}\frac{L^4\gamma^2}{32} Q_0 + O(\gamma^3) \nonumber\\
   &= \left(1 - \frac{9L^4\gamma^2}{32}\right) Q_0 + O(\gamma^3).
 \label{eq:emH}
\end{align}
Crucially, the identity $9\cdot 1^2 - 3^2 = 0$ eliminates all internal wave-packet distortions $B_2(\gamma)$, while $(9\cdot 1^4 - 3^4)/8 = -9$ sets the leading scalar dephasing.

Step 4: Channel reconstruction.
Reassembling the logical blocks yields the recovered joint state
\begin{align}
 \cJ_{\lambda,\gamma}(\rho_L)
 &= \begin{pmatrix} \rho_{00} Q_0 & \rho_{01} H \\ \rho_{10} H & \rho_{11} Q_0 \end{pmatrix} \nonumber\\
 &= \begin{pmatrix} \rho_{00} & (1-2p)\rho_{01} \\ (1-2p)\rho_{10} & \rho_{11} \end{pmatrix} \otimes \rho_Q + O_\diamond(\gamma^3),
\end{align}
with $1 - 2p = 1 - 9L^4\gamma^2/32$, identifying the logical phase-flip probability $p = 9L^4\gamma^2/64$. Incorporating the boundary error $\epsilon_{\rm bd}$ and bounding the finite-dimensional channel remainder in the diamond norm establishes Eq.~\eqref{eq:joint}. Because quadratic envelope distortions cancel identically, the leading memory infidelity depends solely on the lattice spacing $L$, while memory--probe correlations vanish to $O_\diamond(\gamma^3)$.

\textcolor{blue}{\emph{Joint-channel evidence.---}}%
% The reduced-memory infidelity in Fig.~\ref{fig:results}(a) traces out the probe.
% To test whether storage protection and signal extraction coexist in the complete process, we retain the unmeasured probe $Q$ and an entangled reference $R$.
% Let $\cJ_{\lambda,\gamma}$ be the recovered channel with logical output $L$ including its erasure flag, and define its normalized Choi state by
While Fig.~\ref{fig:results}(a) characterizes the reduced memory after tracing out the probe, verifying that storage protection and signal extraction coexist as a compatible instrument requires analyzing the joint output of the recovered memory $L$ (including its erasure flag) and the unmeasured probe $Q$, alongside an entangled reference $R$.
Denoting the complete recovered channel by $\mathcal{J}_{\lambda,\gamma}$, we evaluate its normalized Choi state
\begin{equation}
 J_{\cJ}=(\id_R\otimes\cJ_{\lambda,\gamma})(\ket\Phi\bra\Phi),
 \quad\ket\Phi=(\ket{00}+\ket{11})/\sqrt2.
 \label{eq:emchoi}
\end{equation}
% The target preserves the logical input and outputs the exact lossy probe state $\rho_Q$ of Eq.~\eqref{eq:readout}.
% The Choi trace distance from this product is
Comparing $J_{\mathcal{J}}$ against the ideal product target $|\Phi\rangle\langle\Phi| \otimes \rho_Q$,
where $\rho_Q$ is the exact lossy probe state of Eq.~\eqref{eq:readout}, via the Choi trace distance
\begin{equation}
 D_{\rm Choi}=\frac12\left\|J_{\cJ}-\ket\Phi\bra\Phi\otimes\rho_Q\right\|_1 \label{eq:emdistance}
\end{equation}
% For the two-dimensional logical input, it gives the conservative channel bound
yields the conservative diamond-norm bound for a two-dimensional logical input:
\begin{equation}
 \frac12\left\|\cJ_{\lambda,\gamma}-\id_L\otimes\rho_Q\right\|_\diamond
 \leq 2D_{\rm Choi}.
 \label{eq:embound}
\end{equation}
where $\mathrm{id}_L$ embeds the logical qubit into the accepted no-erasure subspace.
The bound controls arbitrary inputs, including their reference entanglement, and any subsequent probe measurement.
It therefore tests the two outputs and their correlations together, with the signal retained in the probe rather than compared with a signal-free target.

\begin{figure}[b]
 \includegraphics[width=\columnwidth]{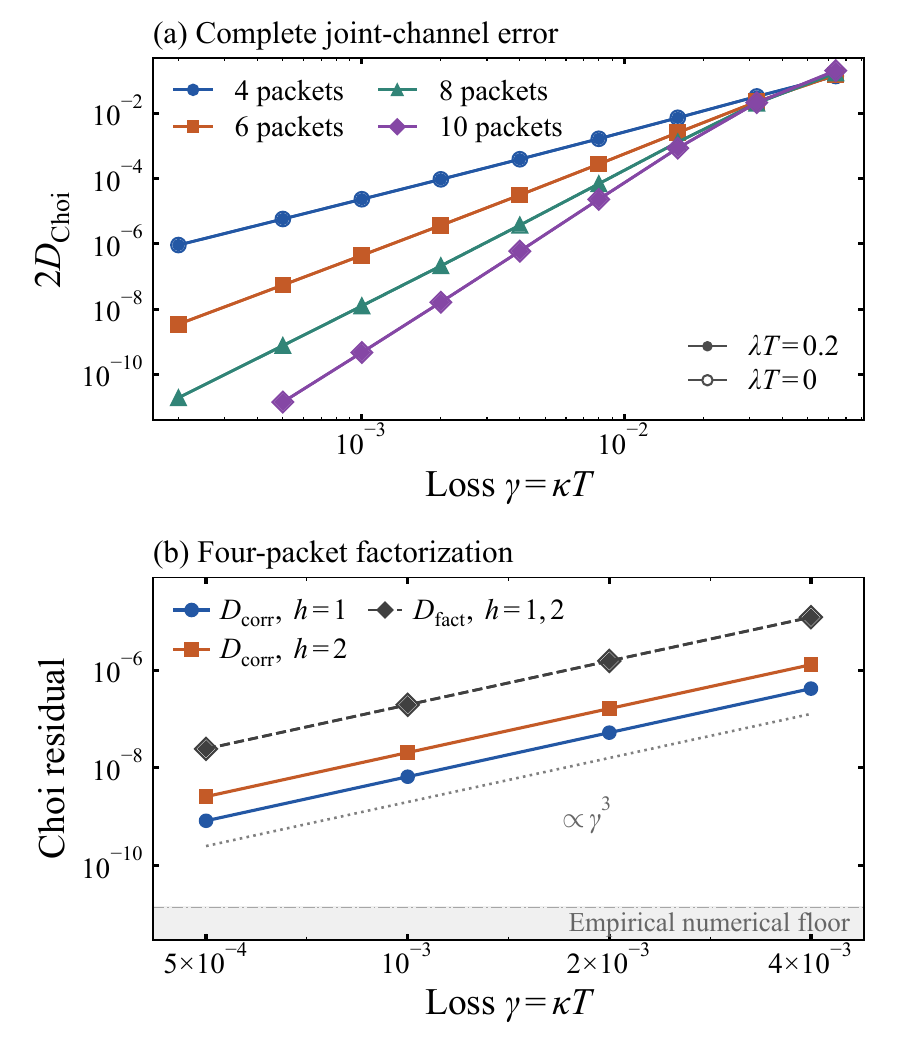}
 \caption{
    Joint-channel tests retaining the reference, erasures, and physical-cell crossings.
    (a) The computed bound $2D_{\rm Choi}$ in Eq.~\eqref{eq:embound} for four through ten packets at $h=1$.
    Filled and open markers show $\lambda T=0.2$ and $0$.
    (b) Four-packet Choi residuals in Eq.~\eqref{eq:emresiduals} at $\lambda T=0.2$ for $h=1,2$.
    Both exhibit cubic scaling in the resolved window; the two $D_{\rm fact}$ curves coincide.
    The dotted guide shows $\gamma^3$; the empirical residual floor is indicated.
    % Unresolved points are omitted in both panels.
    }\label{fig:jointend}
\end{figure}
Figure~\ref{fig:jointend} uses the exact position-space attenuator, calibrated endpoint kick, physical-cell trace, and polynomial site recovery, with all rejected outcomes retained.
For $L=3$, $w=1.8$, and the packet family specified in Fig.~\ref{fig:results}, fits of the joint distance give exponents $2.013$, $3.026$, $4.040$, and $5.081$ for four through ten packets.
The first three fits use $\gamma=0.0002,0.0005,0.001,0.002$; the ten-packet fit omits $0.0002$, where the computed error is below the reliability threshold.
At $\gamma=10^{-3}$ and $\lambda T=0.2$, the respective values of $2D_{\rm Choi}$ are $2.317\times10^{-5}$, $4.422\times10^{-7}$, $1.248\times10^{-8}$, and $4.681\times10^{-10}$, independently confirming the leading error orders predicted by Eqs.~\eqref{eq:order} and \eqref{eq:generaljoint} \cite{supp}.
% \mainrev{These fits apply to the tested codes and loss windows. SM Sec.~V and Table~S2 give convergence checks, numerical floors, and the full hierarchy data~\cite{supp}.}

Because $D_{\mathrm{Choi}}$ in Eq.~\eqref{eq:emdistance} compares against an uncorrupted memory $|\Phi\rangle\langle\Phi|$, the quadratic scaling for four packets in Fig.~\ref{fig:jointend}(a) is dominated by the leading $O(\gamma^2)$ logical dephasing $p = 9L^4\gamma^2/64$. To isolate genuine memory-probe correlations and test the stronger third-order factorization of Eq.~\eqref{eq:joint}, we define the Choi residuals
% \mainrev{To distinguish memory error from residual correlations, the comparison target must be specified.}
% For four packets, Eq.~\eqref{eq:joint} has a third-order remainder relative to the \emph{dephased-memory/exact-probe} product, whose logical dephasing probability is $p=9L^4\gamma^2/64$.
% Equation~\eqref{eq:embound} instead compares with an \emph{unchanged-memory/exact-probe} product and already includes this second-order memory error.
% The quadratic curve in Fig.~\ref{fig:jointend}(a) therefore includes the leading memory error. To isolate nonproduct correlations and test the stronger factorization, define
\begin{align}
 D_{\rm corr}&=\tfrac12\|J_{\cJ}-(\Tr_QJ_{\cJ})\otimes\rho_Q\|_1,\nonumber\\
 D_{\rm fact}&=\tfrac12\|J_{\cJ}-J_{\cD_p}\otimes\rho_Q\|_1,
 \label{eq:emresiduals}
\end{align}
where $J_{\mathcal{D}_p}$ is the Choi state of the analytical leading dephasing channel $\mathcal{D}_p$.
Here, $D_{\mathrm{corr}}$ measures total nonproduct correlations between the memory and probe, while $D_{\mathrm{fact}}$ additionally captures higher-order memory errors beyond $p$.
As shown in Fig. ~\ref{fig:jointend}(b), both residuals clearly exhibit cubic scaling $O(\gamma^3)$ over $\gamma \in [5 \times 10^{-4}, 4 \times 10^{-3}]$ at $h = 1, 2$, with fitted exponents of $3.000$ ($h=1$) and $2.994$ ($h=2$) for $D_{\mathrm{corr}}$, and $2.987$ for $D_{\mathrm{fact}}$, well above the empirical numerical floor.

\end{document}

\title{Supplemental Material for ``Force Sensing in a Recoverable Bosonic Quantum Memory"}

\author{Tairong Bai}
\thanks{These authors contributed equally to this work.}
\affiliation{National Laboratory of Solid State Microstructures, School of Physics, Nanjing University, Nanjing 210093, China}
\affiliation{Shishan Laboratory, Nanjing University, Suzhou 215163, China}
\affiliation{Jiangsu Key Laboratory of Quantum Information Science and Technology, Nanjing University, Suzhou 215163, China}
\author{Wen Zheng}
\thanks{These authors contributed equally to this work}
\email{zhengwen@nju.edu.cn}
\affiliation{National Laboratory of Solid State Microstructures, School of Physics, Nanjing University, Nanjing 210093, China}
\affiliation{Shishan Laboratory, Nanjing University, Suzhou 215163, China}
\affiliation{Jiangsu Key Laboratory of Quantum Information Science and Technology, Nanjing University, Suzhou 215163, China}
\affiliation{ LaunchQ (Suzhou) Technology Co., Ltd., Suzhou 215163, China}
% \author{Xiang Deng}
% \affiliation{National Laboratory of Solid State Microstructures, School of Physics, Nanjing University, Nanjing 210093, China}
% \affiliation{Shishan Laboratory, Nanjing University, Suzhou 215163, China}
% \affiliation{Jiangsu Key Laboratory of Quantum Information Science and Technology, Nanjing University, Suzhou 215163, China}
% \author{Haoyu Zhou}
% \affiliation{National Laboratory of Solid State Microstructures, School of Physics, Nanjing University, Nanjing 210093, China}
% \affiliation{Shishan Laboratory, Nanjing University, Suzhou 215163, China}
% \affiliation{Jiangsu Key Laboratory of Quantum Information Science and Technology, Nanjing University, Suzhou 215163, China}
% \author{Tao Zhang}
% \affiliation{National Laboratory of Solid State Microstructures, School of Physics, Nanjing University, Nanjing 210093, China}
% \affiliation{Shishan Laboratory, Nanjing University, Suzhou 215163, China}
% \affiliation{Jiangsu Key Laboratory of Quantum Information Science and Technology, Nanjing University, Suzhou 215163, China}
\author{Haoyang Cai}
\affiliation{National Laboratory of Solid State Microstructures, School of Physics, Nanjing University, Nanjing 210093, China}
\affiliation{Shishan Laboratory, Nanjing University, Suzhou 215163, China}
\affiliation{Jiangsu Key Laboratory of Quantum Information Science and Technology, Nanjing University, Suzhou 215163, China}
\affiliation{Synergetic Innovation Center of Quantum Information and Quantum Physics, University of Science and Technology of China, Hefei, Anhui 230026, China}
\affiliation{Hefei National Laboratory, Hefei 230088, China}
% \author{Xudong Liao}
% \affiliation{National Laboratory of Solid State Microstructures, School of Physics, Nanjing University, Nanjing 210093, China}
% \affiliation{Shishan Laboratory, Nanjing University, Suzhou 215163, China}
% \affiliation{Jiangsu Key Laboratory of Quantum Information Science and Technology, Nanjing University, Suzhou 215163, China}
\author{Xianke Li}
\affiliation{National Laboratory of Solid State Microstructures, School of Physics, Nanjing University, Nanjing 210093, China}
\affiliation{Shishan Laboratory, Nanjing University, Suzhou 215163, China}
\affiliation{Jiangsu Key Laboratory of Quantum Information Science and Technology, Nanjing University, Suzhou 215163, China}
\author{Yangyang Ge}
\affiliation{National Laboratory of Solid State Microstructures, School of Physics, Nanjing University, Nanjing 210093, China}
\affiliation{Shishan Laboratory, Nanjing University, Suzhou 215163, China}
\affiliation{Jiangsu Key Laboratory of Quantum Information Science and Technology, Nanjing University, Suzhou 215163, China}
\author{Kun Zhou}
\affiliation{National Laboratory of Solid State Microstructures, School of Physics, Nanjing University, Nanjing 210093, China}
\affiliation{Shishan Laboratory, Nanjing University, Suzhou 215163, China}
\affiliation{Jiangsu Key Laboratory of Quantum Information Science and Technology, Nanjing University, Suzhou 215163, China}
\affiliation{ LaunchQ (Suzhou) Technology Co., Ltd., Suzhou 215163, China}
\author{Xiangmin Yu}
\affiliation{National Laboratory of Solid State Microstructures, School of Physics, Nanjing University, Nanjing 210093, China}
\affiliation{Shishan Laboratory, Nanjing University, Suzhou 215163, China}
\affiliation{Jiangsu Key Laboratory of Quantum Information Science and Technology, Nanjing University, Suzhou 215163, China}

\author{Yu Zhang}
\affiliation{National Laboratory of Solid State Microstructures, School of Physics, Nanjing University, Nanjing 210093, China}
\affiliation{Shishan Laboratory, Nanjing University, Suzhou 215163, China}
\affiliation{Jiangsu Key Laboratory of Quantum Information Science and Technology, Nanjing University, Suzhou 215163, China}
\author{Shaoxiong Li}
\affiliation{National Laboratory of Solid State Microstructures, School of Physics, Nanjing University, Nanjing 210093, China}
\affiliation{Shishan Laboratory, Nanjing University, Suzhou 215163, China}
\affiliation{Jiangsu Key Laboratory of Quantum Information Science and Technology, Nanjing University, Suzhou 215163, China}
\affiliation{Synergetic Innovation Center of Quantum Information and Quantum Physics, University of Science and Technology of China, Hefei, Anhui 230026, China}
\affiliation{Hefei National Laboratory, Hefei 230088, China}
\author{Xinsheng Tan}
\affiliation{National Laboratory of Solid State Microstructures, School of Physics, Nanjing University, Nanjing 210093, China}
\affiliation{Shishan Laboratory, Nanjing University, Suzhou 215163, China}
\affiliation{Jiangsu Key Laboratory of Quantum Information Science and Technology, Nanjing University, Suzhou 215163, China}
\affiliation{Synergetic Innovation Center of Quantum Information and Quantum Physics, University of Science and Technology of China, Hefei, Anhui 230026, China}
\affiliation{Hefei National Laboratory, Hefei 230088, China}

\author{Jie Zhao}
\affiliation{National Laboratory of Solid State Microstructures, School of Physics, Nanjing University, Nanjing 210093, China}
\affiliation{Shishan Laboratory, Nanjing University, Suzhou 215163, China}
\affiliation{Jiangsu Key Laboratory of Quantum Information Science and Technology, Nanjing University, Suzhou 215163, China}
\author{Yang Yu}
\email{yuyang@nju.edu.cn}
\affiliation{National Laboratory of Solid State Microstructures, School of Physics, Nanjing University, Nanjing 210093, China}
\affiliation{Shishan Laboratory, Nanjing University, Suzhou 215163, China}
\affiliation{Jiangsu Key Laboratory of Quantum Information Science and Technology, Nanjing University, Suzhou 215163, China}
\affiliation{ LaunchQ (Suzhou) Technology Co., Ltd., Suzhou 215163, China}
\affiliation{Synergetic Innovation Center of Quantum Information and Quantum Physics, University of Science and Technology of China, Hefei, Anhui 230026, China}
\affiliation{Hefei National Laboratory, Hefei 230088, China}

\date{\today}

\maketitle

% ~\\
% ~\\
% ~\\

\tableofcontents

~\\
% ~\\

This Supplemental Material provides the complete mathematical proofs, device benchmark models, and numerical convergence checks supporting the main text.
Section ~\ref{sec:setup} defines the finite-energy wave-packet hierarchy,
derives the exact probe readout at arbitrary integer harmonics,
and specifies the coherent decoding dilation along with its boundary-error bounds.
Building on this framework, Sec.~\ref{sec:kernel} derives the probe-valued loss kernel,
elucidates the four-packet error-cancellation mechanism,
proves joint memory-probe factorization at arbitrary correction order,
and analyzes relative readout-kick calibration errors.
Section~\ref{sec:dephasing} examines intrinsic photon-number dephasing,
demonstrating how the first-order dephasing error present in the four-packet code is eliminated for encodings with six or more packets.
% (while noting that the arbitrary-order vacuum-loss factorization theorem does not extend to dephasing noise).
Section~\ref{sec:device} details the superconducting-cavity benchmarks,
contrasting fixed-input sensitivity against profiled sensitivity in the presence of an unknown logical nuisance parameter, and evaluates the error budgets for endpoint controls, thermal transmon excitations, and multi-cycle sensing with a single stored state.
Finally, Sec.~\ref{sec:numerics} summarizes the experimental hardware parameters, and benchmark datasets.
% \clearpage

\section{Wave-packet hierarchy, exact readout, and decoder dilation}
\label{sec:setup}
% We use $[x,p]=i$, $a=(x+ip)/\sqrt2$, $n=a^\dagger a$, and $H=\hbar\lambda p$. For vacuum loss at rate $\kappa$, write $\gamma=\kappa T$, $\eta=e^{-\gamma}$, $\tau_\kappa=2(1-\sqrt\eta)/\kappa$, and $d=\lambda\tau_\kappa$; $\tau_0=T$. Endpoint encoding, readout, and recovery are ideal unless stated otherwise.

Throughout this work, we set $\hbar = 1$ and adopt canonical coordinates satisfying $[x, p] = i$,
with $a = (x + ip)/\sqrt{2}$, $n = a^\dagger a$, and the force Hamiltonian $H = \lambda p$.
Under continuous photon loss at rate $\kappa$ over duration $T$, we define the dimensionless loss parameter $\gamma = \kappa T$,
the energy transmissivity $\eta = e^{-\gamma}$, the effective accumulation time $\tau_\kappa = 2(1 - \sqrt{\eta})/\kappa$ (with $\tau_0 = T$),
and the net position displacement $d = \lambda \tau_\kappa$.
Unless stated otherwise, state preparation, endpoint probe readout, and recovery operations are treated as ideal instantaneous impulses.

\subsection{Code algebra and finite energy}
% For integer $r\geq1$, set $N=2r+1$, $M=N+1$, $c_j=j-N/2$, and $J=\operatorname{diag}(c_j)$. A normalized real even packet $g$, supported on $|y|<w/2$ with $w<L$, defines
For an integer correction order $r \ge 1$, we set $N = 2r + 1$, $M = N + 1$, site centers $c_j = j - N/2$ ($j = 0, \dots, N$),
and the diagonal site operator $J = \mathrm{diag}(c_j)$.
Choosing a real, even, normalized envelope $g(y)$ compactly supported on $\vert{}y\vert{} < w/2$ with $w < L$,
we define the encoding isometry and logical wave functions as
\begin{equation}
 V_{j\mu}=2^{-(N-1)/2}\sqrt{\binom Nj}\,\bm1_{j\equiv\mu\ ({\rm mod}\ 2)},\qquad
 \psi_\mu(x)=\sum_{j=0}^N V_{j\mu}g(x-Lc_j).
 \label{eq:V}
\end{equation}
The parity-resolved binomial amplitudes use the moment-matching structure familiar from binomial bosonic codes~\cite{binomial,albert}.
Here the weights are assigned to spatial wave-packet sites rather than Fock states, while the internal packet coordinate carries the force signal.
% We assume that the zero extension of $g$ belongs to $H^1(\mathbb R)$, so that $\langle p^2\rangle_g$ is finite; the expansions below require the stronger regularity stated where they are used. Disjoint supports give orthogonal sites. The binomial finite-difference identity implies
We assume that the zero extension of $g$ belongs to the Sobolev space $H^1(\mathbb{R})$ so that the kinetic energy $\langle p^2 \rangle_g$ remains finite; the asymptotic expansions derived below invoke stronger regularity conditions as specified in the corresponding sections. Because the individual wave packets have disjoint spatial supports, the site states are mutually orthogonal.
Moreover, the binomial finite-difference identity implies
\begin{equation}
 \sum_{j=0}^N(-1)^j\binom Nj j^k=0\ (0\leq k<N),\qquad
 V^\dagger J^kV=\mu_k I\ (0\leq k<N),\qquad \mu_1=0,\quad \mu_2=N/4.
 \label{eq:moment}
\end{equation}
Here off-diagonal logical matrix elements vanish automatically because the two logical basis states have disjoint parity support. Denoting the physical code-space projector by $P$ and decomposing the position coordinate within each cell as $x = LJ + y$,
the initial mean photon number is state-independent:
\begin{equation}
 PnP=\bar nP,\qquad
 \bar n=\tfrac12\left(NL^2/4+\langle y^2\rangle_g+\langle p^2\rangle_g-1\right).
 \label{eq:energy}
\end{equation}

For our numerical benchmarks, we employ the smooth finite-energy envelope family
\begin{equation}
 g_m(y)=\sqrt{\frac{4^m}{w\binom{2m}{m}}}\cos^m(\pi y/w)\bm1_{|y|<w/2},\qquad
 m=2r+2,\qquad \langle p^2\rangle_{g_m}=\frac{m^2\pi^2}{(2m-1)w^2}.
 \label{eq:packet}
\end{equation}
Increasing the correction order $r$ thus simultaneously increases the number of wave packets and tightens their spatial localization;
the corresponding mean photon numbers are listed in Table ~\ref{tab:hierarchy}.
Within each wave packet, the annihilation operator acts as $a = (LJ + y + \partial_y)/\sqrt{2}$.
Because pairwise products of operators in $\{I, a, a^\dagger\}$ involve powers of $J$ up to at most $J^2$,
the moment-matching conditions in Eq. ~\eqref{eq:moment} ensure that the discrete Knill-Laflamme criterion $P E_b^\dagger E_c P = \alpha_{bc} P$ for single-photon jumps is already satisfied by the four-packet code ($r = 1$)~\cite{kl}.
Our full finite-time analysis below rigorously accounts for continuous no-jump backaction and higher-order wave-packet distortions.

\subsection{Exact readout at any fixed integer harmonic}
The continuous force-and-loss evolution decomposes into a pure-loss channel $\mathcal{A}_\eta$ followed by a spatial displacement, $\mathcal{E}_{\lambda,\eta}(\rho) = X(d)\mathcal{A}_\eta(\rho)X(d)^\dagger$, with $X(d) = e^{-idp}$. Its position characteristic function transforms as
\begin{equation}
 \Tr[\mathcal E_{\lambda,\eta}(\rho)e^{-ikx}]
 =e^{-ikd-(1-\eta)k^2/4}\Tr[\rho e^{-i\sqrt\eta kx}].
 \label{eq:weyl}
\end{equation}
For any fixed integer harmonic $h \ge 1$, we set the loss-calibrated readout wave number to $q_{\eta,h} = h\pi/(\sqrt{\eta}L)$
and define the envelope form factor $c_{g,h} = \int \vert{}g(y)\vert{}^2 e^{-2\pi i h y/L} dy$.
Because all packet centers share an identical modular phase, the code-space projection obeys $P e^{-2\pi i h x/L} P = (-1)^{hN} c_{g,h} P$.
Consequently, a probe qubit initialized in $\vert{}+\rangle_Q$ and coupled to the oscillator
via the conditional kick $U_h = e^{-i q_{\eta,h} x Z_Q}$ acquires the reduced state
\begin{equation}
 \rho_Q^{(h)}=\frac12\begin{pmatrix}1&\ell_\lambda^{(h)}\\\ell_\lambda^{(h)*}&1\end{pmatrix},\quad
 \ell_\lambda^{(h)}=(-1)^{hN}c_{g,h}e^{-(1-\eta)q_{\eta,h}^2-2iq_{\eta,h}d},\quad
 F_Q^{(h)}=4q_{\eta,h}^2\tau_\kappa^2|\ell_\lambda^{(h)}|^2.
 \label{eq:probe}
\end{equation}
% This marginal is exactly independent of every logical input, including inputs entangled with a reference.
Crucially, this probe marginal is strictly independent of the encoded logical state, even when the memory is entangled with an external reference system.

Writing the probe coherence magnitude as $v = \vert{}l_\lambda^{(h)}\vert{}$
and the signal phase as $\phi = 2q_{\eta,h}d$ (up to the known sign $(-1)^{hN}$),
equally sampling the orthogonal Pauli $X$ and $Y$ probe quadratures yields a classical Fisher information of
\begin{equation}
 F_{XY}=\frac{v^2(\partial_\lambda\phi)^2}{2}
 \left[\frac{\sin^2\phi}{1-v^2\cos^2\phi}+\frac{\cos^2\phi}{1-v^2\sin^2\phi}\right]
 \geq\frac{F_Q^{(h)}}2.
 \label{eq:XY}
\end{equation}
With respect to $\lambda$, the readout phase has period $\sqrt{\eta}L/(h\tau_\kappa)$,
meaning that the rescaled displacement $d/\sqrt{\eta}$ is periodic with period $L/h$.
Since the protected decoding-guard interval has total width $L - w$ in $d/\sqrt{\eta}$,
the condition $h(L - w) < L$ guarantees an injective (unambiguous) phase mapping
across the entire guard range. For our benchmark geometry ($L = 3$, $w = 1.8$) at the
second harmonic ($h = 2$), the phase period $L/h = 1.5$ safely exceeds the guard width $L - w = 1.2$.
Employing even higher harmonics ($h \ge 3$) would require a narrower prior range for $\lambda$ and stronger endpoint displacement kicks.
For comparison, a coherent-state probe subject to the same continuous loss achieves an optimal quantum Fisher information of $F_Q^{\mathrm{coh}} = 2\tau_\kappa^2$ (evaluated on the full oscillator output without equal-energy optimization). 

\subsection{Coherent extraction, recovery, and boundary control}
% Let $s_j=\sqrt\eta Lc_j$, $I_j=[s_j-\sqrt\eta L/2,s_j+\sqrt\eta L/2)$, $\Pi_j$ project onto $I_j$, and $\Pi_{\rm out}=I-\sum_{j=0}^N\Pi_j$. Coherent tagging into a label register $A$ and an acceptance flag $F$, followed by recentering, defines
To model the endpoint decoder, let $s_j = \sqrt{\eta}L c_j$ denote the contracted site centers, $I_j = [s_j - \sqrt{\eta}L/2, s_j + \sqrt{\eta}L/2)$ the corresponding decoding cells, $\Pi_j$ the spatial projector onto $I_j$, and $\Pi_{\mathrm{out}} = I - \sum_{j=0}^N \Pi_j$ the outer erasure projector. Coherently transferring the cell index into an ancillary label register $A$ and an erasure flag $F$, followed by cell recentering, defines the extraction isometry
\begin{equation}
 W=\sum_{j=0}^N\ket{j,0}_{AF}\otimes e^{is_jp}\Pi_j
 +\ket{0,1}_{AF}\otimes\Pi_{\rm out},\qquad W^\dagger W=I,\qquad
 [\rho_A^{\rm acc}]_{jk}=\int_{-\sqrt\eta L/2}^{\sqrt\eta L/2}\rho(s_j+z,s_k+z)\,dz.
 \label{eq:extract}
\end{equation}
Here, $\rho_A^{\mathrm{acc}} = \langle 0\vert{}_F \mathrm{Tr}_G(W\rho W^\dagger)\vert{}0\rangle_F$ represents the unnormalized accepted block after tracing out the recentered gauge oscillator $G$,
while the complete decoder channel also retains the flagged erasure branch ($F = 1$).
Notably, the final integral expression in Eq.~\eqref{eq:extract} holds directly for operator-valued kernels (including the probe and reference systems) without assuming state factorization between the oscillator and the label register.
Within each contracted cell, the readout kick factorizes as
% Here $\rho_A^{\rm acc}=\bra0_F\Tr_G(W\rho W^\dagger)\ket0_F$ is the unnormalized accepted block, where $G$ denotes the recentered oscillator. The full decoder also retains the $F=1$ erasure branch. The final equality holds for probe- and reference-valued kernels and does not assume factorization of the oscillator from the label register. The kick factors as
\[
 e^{-ih\pi Z_Q(c_j+y/L)}=(-1)^{hj}e^{ih\pi NZ_Q/2}e^{-ih\pi Z_Qy/L}.
\]
After undoing the deterministic site-dependent phase via a logical $Z_L^h$ gate,
the accepted site register is recovered using the Kraus operators $R_b = E_b^\dagger$,
where the mutually orthogonal syndrome isometries $E_b = p_b(J)V$ are constructed in Sec~\ref{sec:general}.
Their completeness ensures that $U_{\mathrm{rec}} = \sum_{b=0}^r \vert{}b\rangle_S \otimes R_b$ forms a unitary mapping from the $(2r+2)$-dimensional site space onto the recovered logical qubit and syndrome register $S$ (for the four-packet code, $E_0 = V$ and $E_1 = 2JV/\sqrt{3}$). In all physical-channel evaluations, both accepted and flagged erasure outcomes are fully retained. Implementing this ideal decoder requires $\lceil \log_2(2r+2) \rceil$ ancillary label qubits alongside the flag and probe qubits; in practice,
approximating the sharp spatial projectors $\Pi_j$ within finite gate durations requires compiled continuous-variable control pulses~\cite{eickbusch,terhal}.

Provided the displacement satisfies a positive decoding guard $a_* = (L - w)/2 - \vert{}d\vert{}/\sqrt{\eta} > 0$, the probability $p_{\mathrm{cross}}$ that an individual wave packet leaks outside its designated cell due to loss-induced Gaussian broadening is bounded by
% For a positive guard $a_*=(L-w)/2-|d|/\sqrt\eta$, the loss-induced Gaussian broadening gives an individual packet's probability of leaving its original cell as
\begin{equation}
 p_{\rm cross}\leq\operatorname{erfc}\!\left(a_*\sqrt{\frac{\eta}{1-\eta}}\right),\qquad
 \epsilon_{\rm bd}=\min\{2,4\sqrt{Mp_{\rm cross}}\}.
 \label{eq:boundary}
\end{equation}
This bound accounts for both inter-cell misassignments and outer erasures. To bound the discrepancy between the physical decoder and the analytical original-label channel, let $\vert{}\Psi_j(x, e)\rangle_Q$ denote the normalized Stinespring wave function of the joint oscillator-environment-probe system evolved from input packet $j$.
We define an auxiliary virtual isometry $W_v$ that retains the initial site label over the entire real line:
% This includes crossings into neighboring cells, in addition to erasure. To compare with the original-label analytic channel, let $\ket{\Psi_j(x,e)}_Q$ be the normalized loss--probe Stinespring wave function for input packet $j$. Define the auxiliary isometry
\begin{equation}
 W_{\rm v}\ket j=\ket j_A\int_{\mathbb R}dz\,de\,\ket z_G\ket e_E\ket{\Psi_j(z+s_j,e)}_Q.
 \label{eq:virtual}
\end{equation}
Because the site labels are orthogonal, $W_v^\dagger W_v = I_M$,
which defines a valid completely positive and trace-preserving (CPTP) virtual comparison channel $\mathcal{J}_{\mathrm{virt}}$.
Embedding the virtual labels into the flagged space via $\vert{}j\rangle_A \equiv \vert{}j, 0\rangle_{AF}$ places both channels on the identical output Hilbert space.
Applying the physical extraction isometry $W_p \equiv W$ of Eq.~\eqref{eq:extract} to the same Stinespring state yields an output that coincides with $W_v\vert{}j\rangle$ inside the original cell $I_j$, while each of the two complementary tail contributions has norm at most $\sqrt{p_{\mathrm{cross}}}$.
Consequently, $\Vert{}(W_p - W_v)\vert{}j\rangle\Vert{} \le 2\sqrt{p_{\mathrm{cross}}}$, which implies the operator-norm bound $\Vert{}W_p - W_v\Vert{} \le 2\sqrt{M p_{\mathrm{cross}}}$.
Since subsequent phase correction, partial traces, and recovery are trace-norm contractions, the diamond-norm distance between the physical and virtual joint channels satisfies $\Vert{}\mathcal{J}_{\mathrm{phys}} - \mathcal{J}_{\mathrm{virt}}\Vert{}_\diamond \le \epsilon_{\mathrm{bd}}$ (including arbitrary reference entanglement).
By construction, $\mathcal{J}_{\mathrm{virt}}$ tracks the original packet labels analytically rather than extracting them spatially;
for any fixed positive guard $a_* > 0$, the boundary discrepancy $\epsilon_{\mathrm{bd}}$ vanishes exponentially as $\exp(-b/\gamma)$.

\section{Probe-valued kernel and joint-channel factorization}
\label{sec:kernel}
Tracing out the vacuum environment port of an equivalent beam-splitter transformation yields the exact finite-loss position-space kernel:
\begin{align}
 \rho_{\rm out}(x,x')&=\frac{e^{-(1-\eta)(x-x')^2/4}}{\sqrt{\pi\eta}}
 \int d\xi\,e^{-\xi^2}\rho_{\rm in}(u_+,u_-),\label{eq:poskernel}\\
 u_\pm&=\frac{(x+x')/2-d}{\sqrt\eta}\pm\frac{\sqrt\eta(x-x')}{2}
 -\sqrt{\frac{1-\eta}{\eta}}\xi.\nonumber
\end{align}
Our exact physical-channel simulations evaluate Eq.~\eqref{eq:extract} by integrating the full encoded wave functions at $u_\pm$ over the finite cell intervals $I_j$. For the analytical virtual channel $\mathcal{J}_{\mathrm{virt}}$, we extend the recentered coordinate integral to $\mathbb{R}$ while keeping track of the original site indices.
To express the resulting site couplings, we define the envelope autocorrelation and shifted Fourier-overlap integrals:
\begin{equation}
 A_g(s)=\int g(y-s/2)g(y+s/2)dy,\qquad
 C_{g,h}(s)=\int g(y-s/2)g(y+s/2)e^{-2\pi ihy/L}dy.
 \label{eq:overlap}
\end{equation}
Because $g(y)$ is real and even, both $A_g(s)$ and $C_{g,h}(s)$ are real, even functions of the drift shift $s$. Setting the site separation $m = j - k$, relative drift $s_m = (1 - \eta)Lm$, inter-site Gaussian damping factor $u = \eta(1 - \eta)L^2/4$, probe dephasing exponent $B_h = (1 - \eta)q_{\eta,h}^2$, and signal phase $\phi_h = 2q_{\eta,h}d$, the probe-valued site kernel after undoing the deterministic logical phase $Z_L^h$ takes the form
\begin{equation}
 \begin{gathered}
 Q_m^{(h)}=\frac{e^{-um^2}}2
 \begin{pmatrix}
 A_g(s_m)&(-1)^{hN}e^{-B_h-i\phi_h}C_{g,h}(s_m)\\
 (-1)^{hN}e^{-B_h+i\phi_h}C_{g,h}(s_m)&A_g(s_m)
 \end{pmatrix},\\
 Q_0^{(h)}=\rho_Q^{(h)},\quad K_m=\Tr Q_m^{(h)}=e^{-um^2}A_g(s_m).
 \end{gathered}
 \label{eq:harmonickernel}
\end{equation}
Notably, tracing out the probe yields a reduced memory kernel $K_m = \mathrm{Tr}\,Q_m^{(h)}$ that is strictly independent of the readout harmonic $h$.  

\subsection{Four-packet cancellation mechanism}
For the four-packet code ($r = 1$), applying the recovery operators preserves the diagonal logical blocks as $\rho_{\mu\mu}|\mu\rangle\langle\mu| \otimes Q_0^{(h)}$ while mapping the off-diagonal blocks to $\rho_{01}|0\rangle\langle 1| \otimes H + \mathrm{H.c.}$, where the effective off-diagonal probe kernel and resulting entanglement fidelity are given by
% For $r=1$, recovery leaves diagonal logical blocks $\rho_{\mu\mu}\ket\mu\bra\mu\otimes Q_0^{(h)}$ and off-diagonal blocks $\rho_{01}\ket0\bra1\otimes H+\mathrm{H.c.}$, where
\begin{equation}
 H=\frac{9Q_1^{(h)}-Q_3^{(h)}}8,\qquad
 F_e=\frac12+\frac{9K_1-K_3}{16}.
 \label{eq:fourkernel}
\end{equation}
Specifically, across the four opposite-parity site pairs, the recovery algebra assigns weights of $+3/8$ to each of the three $|m| = 1$ pairs (summing to $9/8$) and $-1/8$ to the $|m| = 3$ pair. Expanding the overlap integrals for a sufficiently smooth envelope yields $A_g(s) = 1 - \langle p^2 \rangle_g s^2/2 + O(s^4)$ and $C_{g,h}(s) = c_{g,h} + O(s^2)$, which leads to the small-loss expansion at fixed $h$:
\begin{equation}
 Q_m^{(h)}=Q_0^{(h)}+m^2B_{2,h}(\gamma)
 +\frac{L^4\gamma^2}{32}m^4Q_0^{(h)}+O(\gamma^3),\qquad
 H=\left(1-\frac{9L^4\gamma^2}{32}\right)Q_0^{(h)}+O(\gamma^3).
 \label{eq:fourcancel}
\end{equation}
Here, the operator $B_{2,h}(\gamma) = O(\gamma)$ collects all envelope-dependent wave-packet distortions through second order in $\gamma$.
Crucially, because $9 \cdot 1^2 - 3^2 = 0$, the linear combination in $H$ identically cancels the entire $m^2 B_{2,h}(\gamma)$ contribution, leaving only the scalar $m^4$ term proportional to $Q_0^{(h)}$. Consequently, the recovered joint channel factorizes as $\mathcal{D}_p \otimes \rho_Q^{(h)} + O_\diamond(\gamma^3) + O_\diamond(\epsilon_{\mathrm{bd}})$, where $\mathcal{D}_p(\rho) = (1 - p)\rho + p Z_L \rho Z_L$ is a pure logical dephasing channel with phase-flip probability $p = 9L^4\gamma^2/64$. While comparing the joint output directly against an uncorrupted identity memory channel yields a leading distance of $O(\gamma^2)$ due to $p$, isolating the factorized target $\mathcal{D}_p \otimes \rho_Q^{(h)}$ reveals that genuine memory-probe correlations vanish to $O_\diamond(\gamma^3)$. 

\subsection{Arbitrary-order factorization and memory scaling}
\label{concise:arbitraryjoint}\label{sec:general}
We now generalize this factorization theorem to arbitrary correction order $r \ge 1$, with $N = 2r + 1$, $L > w > 0$, and any fixed integer harmonic $h \ge 1$.
We assume a real, even, normalized envelope $g(y)$ supported on $[-w/2, w/2]$ whose zero extension belongs to $H^{r+2}(\mathbb{R})$,
and we require the decoding guard $a_*$ to remain bounded below by a positive constant as $\gamma \to 0$.
All asymptotic bounds below hold at fixed $r, L, g,$ and $h$ (the implicit constants need not be uniform in $r$ or $h$).

By virtue of Eq.~\eqref{eq:moment}, the candidate syndrome subspaces spanned by $\{V, JV, \dots, J^r V\}$ share an identical Gram matrix across both logical parity sectors.
Gram-Schmidt orthogonalization therefore generates a single set of real polynomials $p_b(J)$ ($0 \le b \le r$) that define mutually orthogonal syndrome isometries $E_b = p_b(J)V$.
Denoting the binomial site weights by $w_j = \binom{N}{j}/2^{N-1}$, each parity sector comprises $r + 1$ sites, meaning that the $(r+1) \times (r+1)$ matrix with entries $\sqrt{w_j}p_b(c_j)$ is orthogonal.
This guarantees completeness of the recovery operators $R_b = E_b^\dagger$ ($\sum_{b=0}^r R_b^\dagger R_b = I_{2r+2}$). When applied to the virtual channel kernels in Eq.~\eqref{eq:harmonickernel}, completeness preserves the diagonal logical blocks as $\rho_{\mu\mu}|\mu\rangle\langle\mu| \otimes Q_0$ (with $Q_0 \equiv \rho_Q^{(h)}$).
Furthermore, because $g(y)$ is real and even, the kernels satisfy $Q_{-m} = Q_m = Q_m^\dagger$, allowing us to express the virtual recovered channel as
\begin{align}
 \widetilde{\cJ}_r(\rho)
 &=\sum_{\mu=0}^1\rho_{\mu\mu}\ket\mu\bra\mu\otimes Q_0
 +\rho_{01}\ket0\bra1\otimes H_r
 +\rho_{10}\ket1\bra0\otimes H_r,\nonumber\\
 H_r&=\sum_{k=0}^r c_{r,k}Q_{2k+1}.
 \label{concise:jointclosed}
\end{align}
% Before grouping separations, an even--odd pair $(j,l)$ has coefficient $w_jw_l\sum_b p_b(c_j)p_b(c_l)$.
Prior to grouping terms by site separation $m = |j - l|$, each opposite-parity site pair $(j, l)$ contributes with weight $w_j w_l \sum_{b=0}^r p_b(c_j)p_b(c_l)$. 

To determine the grouped recovery coefficients $c_{r,k}$ explicitly, we evaluate the action of the same recovery map on an auxiliary unitary site rotation $e^{-i\theta J}$. In the logical basis, the recovered Kraus operators are diagonal, $\mathrm{diag}(f_{b0}, f_{b1})$, yielding the off-diagonal coherence factor $\chi(\theta) = \sum_{b=0}^r f_{b0} f_{b1}^*$.
Trace preservation implies
% To identify these coefficients, apply the same recovery to the auxiliary rotation $e^{-i\theta J}$. Write its diagonal Kraus operators as $\operatorname{diag}(f_{b0},f_{b1})$ and its coherence factor as $\chi=\sum_bf_{b0}f_{b1}^*$. Trace preservation gives
\begin{equation}
 1-\operatorname{Re}\chi=\tfrac12\sum_b|f_{b0}-f_{b1}|^2
 =O(\theta^{2r+2}).
 \label{concise:rotation}
\end{equation}
Because $p_b(J)J^q$ is a polynomial in $J$ of degree at most $2r$ for all $q \le r$, the moment-matching conditions in Eq.~\eqref{eq:moment} ensure that the Taylor expansions of $f_{b0}(\theta)$ and $f_{b1}(\theta)$ coincide through order $\theta^r$, which establishes the $O(\theta^{2r+2})$ scaling in Eq.~\eqref{concise:rotation}.
On the other hand, grouping the site pairs by odd separations $2k + 1$ gives $\mathrm{Re}\,\chi(\theta) = \sum_{k=0}^r c_{r,k} \cos[(2k+1)\theta]$. Matching the Taylor series of both expressions around $\theta = 0$ uniquely identifies $c_{r,k}$ as the Lagrange interpolation weights at zero for the nodes $t_k = (2k+1)^2$:
\begin{equation}
 c_{r,k}=\prod_{l\ne k}\frac{-t_l}{t_k-t_l},\qquad
 \sum_kc_{r,k}t_k^n=
 \begin{cases}1,&n=0,\\0,&1\leq n\leq r,\\(-1)^r[(2r+1)!!]^2,&n=r+1.\end{cases}
 \label{concise:weights}
\end{equation}
% The last identity follows by evaluating the interpolation remainder for $t^{r+1}$ at zero.
where the final identity follows from the standard Lagrange remainder theorem applied to the monomial $t^{r+1}$ at $t = 0$.

% The reduced site kernel also admits an exact random-rotation representation,
The reduced memory kernel also admits an exact representation as a classical mixture of random site rotations:
\begin{equation}
 \operatorname{Tr}Q_m=\mathbb E[e^{-im\theta}],\qquad
 \theta=L\sqrt{\eta(1-\eta)/2}\,Z_0+(1-\eta)LP_g,
 \label{concise:phase}
\end{equation}
where $Z_0$ is a standard normal random variable and $P_g$ is an independent random momentum variable distributed according to the normalized Fourier density $|\tilde{g}(p)|^2$.
% Viewed in the symmetric Dicke subspace of $N$ formal spin-$1/2$ particles, the polynomial syndrome subspaces spanned by $p_b(J)V$ correspond precisely to the symmetric phase-flip error spaces of weight $b \le r$.
Using the symmetric-spin representation associated with the binomial amplitudes~\cite{binomial,albert}, the polynomial syndrome subspaces spanned by $p_b(J)V$ correspond precisely to the symmetric phase-flip error spaces of weight $b\le r$ in the Dicke subspace of $N$ formal spin-$1/2$ particles.
Majority-vote decoding therefore yields the exact conditional entanglement infidelity for a fixed rotation angle $\theta$:
\begin{equation}
 p_r(\theta)=\sum_{b=r+1}^{N}\binom Nb
 \sin^{2b}(\theta/2)\cos^{2(N-b)}(\theta/2).
 \label{concise:majority}
\end{equation}
Because the probability distribution of $\theta$ in Eq.~\eqref{concise:phase} is symmetric,
any coherent logical Z rotation averages to zero.
Taking the expectation value of Eq.~\eqref{concise:majority} over the leading Gaussian fluctuation in $\theta$
yields the asymptotic memory infidelity $\bar{p}_r = \mathbb{E}[p_r(\theta)] = a_r \gamma^{r+1} + O(\gamma^{r+2})$, with the prefactor $a_r$ given explicitly below.  

Establishing joint memory-probe factorization requires extending this scalar cancellation to the matrix-valued kernels $Q_m$.
Setting $\delta = 1 - \eta$ and $u = \eta\delta L^2/4$, we write Eq.~\eqref{eq:harmonickernel} as $Q_m = e^{-um^2} T_\gamma(\delta Lm)$,
where $T_\gamma(s)$ is an even matrix-valued function satisfying $T_\gamma(0) = Q_0$.
Under our assumption that $g \in H^{r+2}(\mathbb{R})$, the Cauchy-Schwarz inequality guarantees that both the autocorrelation $A_g(s)$
and the shifted Fourier overlap $C_{g,h}(s) = \int dp\,\tilde{g}(p - k_h/2)^* \tilde{g}(p + k_h/2) e^{ips}$ (with $k_h = 2\pi h/L$) possess bounded continuous derivatives through order $2r + 4$.
Expanding $T_\gamma(s)$ in a Taylor series with trace-norm remainder $O_1(s^{2r+2})$ and applying the moment identities of Eq.~\eqref{concise:weights} yields
\begin{align}
 T_\gamma(s)&=Q_0+\sum_{a=1}^rs^{2a}T_{2a}(\gamma)+O_1(s^{2r+2}),\nonumber\\
 \delta^{2a}\sum_kc_{r,k}t_k^ae^{-ut_k}
 &=O(\delta^{2a}u^{r+1-a})=O(\gamma^{r+1+a}),\quad 1\leq a\leq r.
 \label{concise:matrixcancel}
\end{align}
where the expansion coefficients and trace-norm remainder bounds are uniform for small $\gamma$.
Specifically, the algebraic identities in Eq.~\eqref{concise:weights} annihilate every cross term of combined order $1 \le a + b \le r$
arising from the $a$-th power of the packet-drift expansion and the $b$-th power of the Gaussian damping expansion $e^{-ut_k}$.
Moreover, the Taylor remainder of the envelope overlap scales as $O(\delta^{2r+2})$, which is at most $O(\gamma^{r+2})$ for all $r \ge 1$.
Consequently, all envelope-dependent probe distortions vanish through order $\gamma^{r+1}$.
Substituting $u = L^2\gamma/4 + O(\gamma^2)$ into the surviving $n = r + 1$ Gaussian term of Eq.~\eqref{concise:weights} gives
\begin{equation}
 H_r=(1-2\pi_r)Q_0+O_1(\gamma^{r+2}),\qquad
 \pi_r=a_r\gamma^{r+1},\qquad
 a_r=\frac{\binom{2r+1}{r+1}(2r+1)!!}{8^{r+1}}L^{2r+2}.
 \label{concise:hierarchy}
\end{equation}

Finally, comparing Eq.~\eqref{concise:jointclosed} against the factorized target channel $\mathcal{D}_{\pi_r} \otimes Q_0$,
where $\mathcal{D}_\pi(\rho) = (1 - \pi)\rho + \pi Z\rho Z$,
shows that the two maps coincide on the diagonal logical blocks
and differ only on the off-diagonal blocks. For any normalized input state $\rho$ (including arbitrary reference entanglement),
the off-diagonal block satisfies $\|\rho_{01}\|_1 \le 1/2$, a bound saturated by the logical $|+\rangle\langle +|$ state.
We thus obtain the diamond-norm bounds
\begin{align}
 \tfrac12\|\widetilde{\cJ}_r-\mathcal D_{\pi_r}\otimes Q_0\|_\diamond
 &=\tfrac12\|H_r-(1-2\pi_r)Q_0\|_1=O(\gamma^{r+2}),\nonumber\\
 \|\cJ_r-\mathcal D_{\pi_r}\otimes\rho_Q^{(h)}\|_\diamond
 &\leq C\gamma^{r+2}+\epsilon_{\rm bd}.
 \label{concise:generalbound}
\end{align}
Combining this virtual-channel result with the boundary-crossing bound of Eq.~\eqref{eq:boundary} for $M = 2r + 2$ completes the proof for the physical joint channel $\mathcal{J}_r$, retaining all cell crossings, flagged erasures, and reference correlations. (Note that replacing the leading-order dephasing probability $\pi_r$ with the exact reduced-memory infidelity $\bar{p}_r = (1 - \mathrm{Tr}\,H_r)/2$ leaves the $O(\gamma^{r+2})$ remainder scaling unchanged.) While our numerical $\cos^{2r+2}$ wave packets satisfy the required $H^{r+2}(\mathbb{R})$ regularity, residual Kerr nonlinearity, photon-number dephasing, and finite-duration control errors lie outside the scope of this pure-loss factorization theorem.

\subsection{Relative kick calibration}
% For $q_{\rm act}=(1+\varepsilon)h\pi/(\sqrt\eta L)$, the parity-binomial sum gives the exact basis-readout difference
If the conditional readout kick experiences a relative wave-number calibration error $\epsilon$,
such that $q_{\mathrm{act}} = (1 + \epsilon)h\pi/(\sqrt{\eta}L)$,
evaluating the parity-alternating binomial sum yields an exact bound on the probe readout discrepancy between the two logical basis states:
\begin{equation}
 |\ell_0-\ell_1|=2|c_{g,h}(\varepsilon)|e^{-(1-\eta)q_{\rm act}^2}
 |\sin(\pi h\varepsilon)|^{2r+1},\qquad
 c_{g,h}(\varepsilon)=\int |g(y)|^2e^{-2\pi ih(1+\varepsilon)y/L}dy.
 \label{eq:kickerror}
\end{equation}
Because the logical basis states have disjoint site support, off-diagonal logical matrix elements continue to vanish identically.
Thus, at any fixed harmonic $h$, logical-state contamination of the probe readout is suppressed to $O(|\epsilon|^{2r+1})$ in the relative kick error.
Note that this quantifies state-dependent readout variation rather than systematic force-estimation bias or decoder miscalibration.  

\section{Intrinsic number dephasing and first-order correction}
\label{sec:dephasing}
% Write $\delta_\phi=T/T_\phi$ for the generator $2\mathcal D[n]/T_\phi$, distinct from $\delta=1-\eta$ used in the loss-kernel proof. For the memory-error expansions in this section, assume that the zero extension of $g$ belongs to $H^4(\mathbb R)$; all numerical packets in Eq.~\eqref{eq:packet} satisfy this sufficient condition. In site--gauge coordinates,
% $n=L^2J^2/2+LJy+B_g$, where $B_g=(y^2+p_y^2-1)/2$. For four packets the loss-recovery syndromes give

To analyze intrinsic photon-number dephasing generated by $(2/T_\phi)\mathcal{D}[n]$,
we define the dimensionless dephasing parameter $\delta_\phi = T/T_\phi$, distinct from $\delta = 1 - \eta$ used in Sec. \ref{sec:kernel}.
For the perturbative memory-infidelity expansions below, we assume that the zero extension of $g(y)$ belongs to $H^4(\mathbb{R})$,
a condition satisfied by all envelopes in Eq.~\eqref{eq:packet}. In terms of the site-gauge decomposition $x = LJ + y$,
the photon-number operator expands as $n = L^2 J^2/2 + LJy + B_g$, where $B_g = (y^2 + p_y^2 - 1)/2$ acts purely on the internal coordinate.
For the four-packet code ($r = 1$), projecting $n$ onto the loss-recovery syndrome spaces yields
\begin{equation}
 R_0nV=I_L\otimes(B_g+3L^2/8),\qquad
 R_1nV=-\frac{\sqrt3L^2}{4}Z_L\otimes I_g+I_L\otimes\frac{\sqrt3L}{2}y,
 \qquad (1-F_e)_\phi=\frac{3L^4}{8}\delta_\phi+O(\delta_\phi^2).
 \label{eq:fourphase}
\end{equation}
% The first-order error is the logical $Z_L$ component aliased into the $JV$ syndrome. Six packets separate it: with $\mu_2=5/4$, $\mu_4=65/16$, use
Here, the first-order memory infidelity arises because the $J^2$ component of $n$ aliases into the $JV$ syndrome subspace as an uncorrectable logical $Z_L$ error.
Encoding into six packets ($r = 2$) resolves this aliasing, that is, using the moments $\mu_2 = 5/4$ and $\mu_4 = 65/16$, we construct three mutually orthogonal syndrome isometries:
\begin{align}
 E_0&=V,\quad E_1=2JV/\sqrt5,\quad
 E_2=(J^2-5/4)V/\sqrt{5/2},\qquad E_b^\dagger E_c=\delta_{bc}I,\nonumber\\
 n(V\otimes I_g)&=\sum_{b=0}^2E_b\otimes G_b,\quad
 G_0=B_g+5L^2/8,\quad G_1=\sqrt5Ly/2,\quad
 G_2=\tfrac{L^2}{2}\sqrt{5/2}\,I_g.
 \label{eq:sixnumber}
\end{align}
Because each gauge operator $G_b$ acts exclusively on the internal coordinate $y$,
the standard binomial recovery $R_b = E_b^\dagger$ completely corrects the first-order dephasing jump $n(V \otimes I_g)$,
while the identity $V^\dagger n^2 V = I_L \otimes \sum_{b=0}^2 G_b^\dagger G_b$ cancels the accompanying first-order no-jump backaction.
Tracing out the internal gauge coordinate therefore eliminates the entire $O(\delta_\phi)$ memory error.
Because encodings with $r \ge 2$ match site moments through at least degree four, this first-order cancellation holds for all codes with six or more packets;
moreover, a bounded force displacement shifts only the shared internal coordinate and leaves this site-moment orthogonality intact.
Retaining the regularity assumptions of Sec.~\ref{sec:general} at a fixed positive guard (in the absence of Kerr nonlinearity), the combined reduced-memory infidelity scales as
% Every $G_b$ acts only on the gauge.
% The same recovery $R_b=E_b^\dagger$ corrects the complete first-order jump, while $V^\dagger n^2V=I_L\otimes\sum_bG_b^\dagger G_b$ supplies the no-jump normalization.
% The recovered first-order generator is gauge-only and vanishes after its partial trace. Matching moments through degree four extends this conclusion to every $r\geq2$;
% a bounded force translation changes only the common gauge coordinate. For the combined expansion, retain the vacuum-loss regularity assumptions of Sec.~\ref{sec:general} as well. At fixed positive guard and with Kerr absent,
\begin{equation}
 (1-F_e)_\phi=O(\delta_\phi^2),\qquad
 1-F_e=O(\gamma^{r+1})+O(\delta_\phi^2)+O(\gamma\delta_\phi),\quad r\geq2.
 \label{eq:mixednoise}
\end{equation}
% This is a logical-memory statement. It neither establishes arbitrary-order number-dephasing suppression nor extends the vacuum-loss joint-instrument theorem to Kerr or number dephasing. Equation~\eqref{eq:boundary} bounds vacuum-loss crossings; phase-noise boundary effects are retained in the physical-cell calculation.
We emphasize that Eq.~\eqref{eq:mixednoise} characterizes the reduced logical memory:
it neither implies arbitrary-order suppression of number dephasing nor extends the joint memory-probe factorization theorem of Sec.~\ref{sec:general} to dephasing or Kerr channels.
While Eq.~\eqref{eq:boundary} bounds vacuum-loss cell crossings, any phase-noise-induced boundary leakage is fully captured in our numerical physical-cell evaluations.

To verify this scaling transition independently, we simulate pure number dephasing in the Fock basis via the exact element-wise attenuation $\rho_{nm} \mapsto e^{-\delta_\phi (n-m)^2}\rho_{nm}$ (setting $\lambda = \kappa = K = 0$).
At $L = 3$ and $w = 1.8$, log-log fits over $\delta_\phi \in \{2 \times 10^{-6}, 5 \times 10^{-6}, 10^{-5}, 2 \times 10^{-5}\}$ yield scaling exponents of $0.99966$ for the four-packet code and $1.99707$ for the six-packet code.
At $\delta_\phi = 10^{-5}$, the corresponding memory infidelities are $3.03619 \times 10^{-4}$ and $1.41070 \times 10^{-6}$, respectively, well above the numerical zero-noise floors of $1.05 \times 10^{-9}$ and $7.87 \times 10^{-11}$.

\section{Superconducting-cavity benchmarks, readout diagnostics, and control costs}
\label{sec:device}

Our hardware benchmarks adopt experimental parameters from a state-of-the-art three-dimensional superconducting cavity-transmon system~\cite{milul}, as summarized in Table~\ref{tab:device}.
During the sensing interval, the oscillator evolves under the combined master equation
\begin{equation}
 \dot\rho=-i[\lambda p-Kn(n-1)/2,\rho]+\frac{\mathcal D[a]\rho}{T_1^c}
 +\frac{2\mathcal D[n]\rho}{T_\phi}.
 \label{eq:deviceME}
\end{equation}
We compare four physical noise models:
(i) pure photon loss;
(ii) photon loss plus residual Kerr nonlinearity;
(iii) the combined "stress" model incorporating loss, Kerr nonlinearity,
and a conservative lower-bound cavity dephasing time $T_\phi = 0.5\text{ s}$;
and (iv) the combined stress evolution followed by an ideal signal-independent endpoint Kerr-reversal unitary $U_{\mathrm{inv}} = e^{-iKTn(n-1)/2}$.
While $U_{\mathrm{inv}}$ exactly removes isolated Kerr phase accumulation, our full time-domain propagation retains the noncommutation of Kerr nonlinearity with continuous force displacement and photon loss.
Finite-duration endpoint gate imperfections and parasitic transmon thermal jumps are budgeted separately below.   

\subsection{Conditional sensitivity and logical nuisance} \label{sec:sensitivity}
\begin{figure}[t]
\centering
\includegraphics[width=0.7\textwidth]{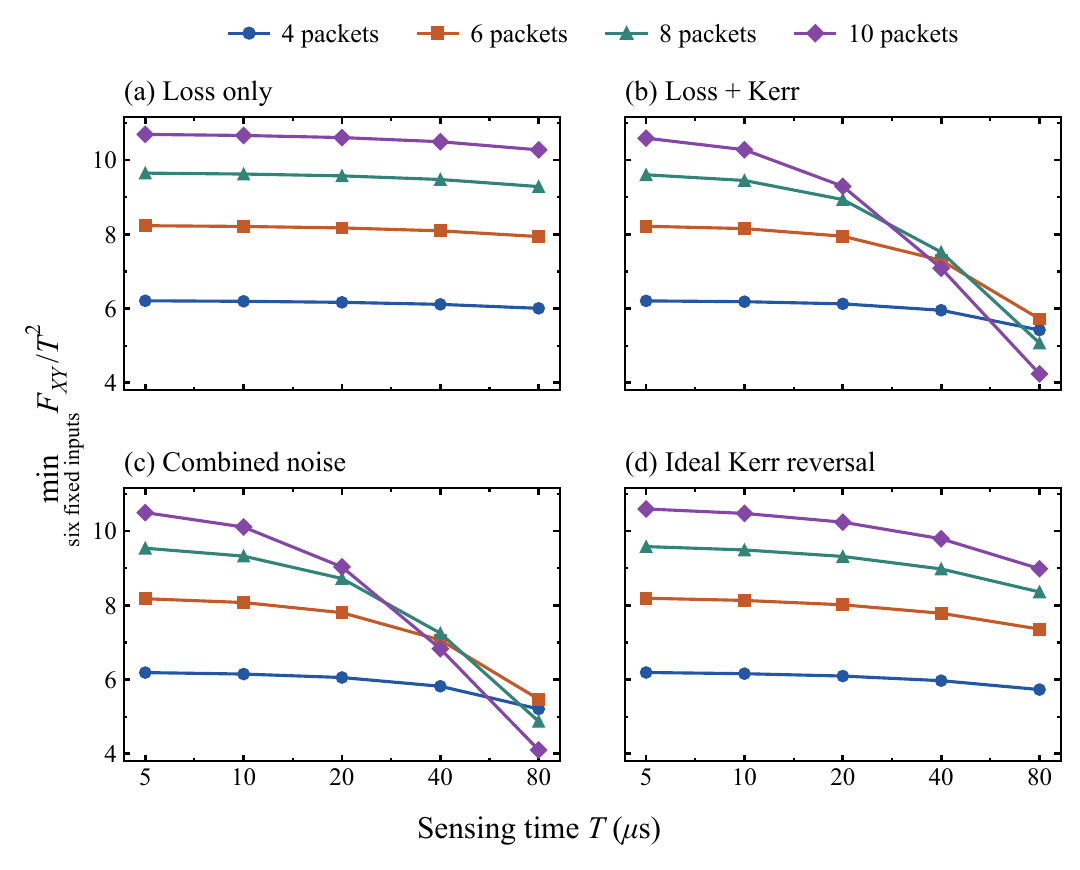}
\caption{
    Worst-case conditional probe sensitivity complementing the memory-infidelity benchmarks in Fig. 3 of the main text.
    Panels show the minimum normalized classical Fisher information $\min F_{XY}/T^2$ across the six logical Pauli eigenstates versus sensing duration $T$
    under (a) pure photon loss,
    (b) photon loss and residual Kerr nonlinearity,
    (c) combined photon loss, Kerr nonlinearity, and intrinsic number-dephasing stress ($T_\phi = 0.5\text{ s}$),
    and (d) the combined stress evolution followed by ideal endpoint Kerr reversal.
    All curves are evaluated for four through ten wave packets at $L = 3$, $w = 1.8$, $h = 2$,
    and fixed dimensionless impulse $\lambda T = 0.2$ (with the logical input state held fixed when evaluating signal derivatives).
    Endpoint controls are treated as ideal and parasitic transmon heating events are excluded. 
    % Conditional readout complementing the memory errors in main-text Fig.~3. Panels show (a) loss only, (b) loss with Kerr, (c) loss, Kerr, and intrinsic-dephasing stress, and (d) the same stress followed by ideal endpoint Kerr reversal. Colors and markers identify four through ten packets; lines guide the eye. All panels use $L=3$, $w=1.8$, $h=2$, and $\lambda T=0.2$ at each duration. The ordinate is the minimum equal-$X/Y$ Fisher information over six fixed logical Pauli eigenstates, each held fixed in the signal derivative. It is a conditional benchmark; it does not minimize over every logical input or profile an unknown logical state. Endpoint operations are ideal; thermal transmon events are excluded.
    }
\label{fig:conditionalsensitivity}
\end{figure}

% For the finite-basis isometry $V_F$ and the selected accumulation/reversal channel $\mathcal E$, define
Denoting the Fock-truncated encoding isometry by $V_F$ and the selected accumulation-and-reversal channel by $\mathcal{E}$, we characterize logical-state-dependent readout perturbations via the $2 \times 2$ logical probe matrix
\begin{equation}
 B_{ji}=\Tr[\mathcal E(V_F\ket i\bra jV_F^\dagger)e^{-2iq_{\eta,h}x}],\qquad
 \delta_{\rm read}=\|B-\tfrac12\Tr(B)I\|_\infty,\qquad
 |\ell(\rho_L)-\ell(I/2)|\leq\delta_{\rm read}.
 \label{eq:dependence}
\end{equation}
where $\|\cdot\|_\infty$ denotes the operator spectral norm (largest singular value),
so that $\delta_{\mathrm{read}}$ bounds the maximum deviation of the probe coherence $l(\rho_L)$ from its maximally mixed value $l(I/2)$.
Parameterizing the dimensionless impulse as $\zeta = \lambda T$ at each fixed duration $T$ and writing $u = \mathrm{Re}\,l$ and $v = \mathrm{Im}\,l$,
the normalized classical Fisher information for equal $X/Y$ Ramsey sampling with a fixed known logical input is
% Here $\|\cdot\|_\infty$ is the largest singular value. At each fixed duration let $\zeta=\lambda T$, $u=\Re\ell$, and $v=\Im\ell$. The plotted conditional information is
\begin{equation}
 \frac{F_{XY}}{T^2}=\frac12\left[\frac{(\partial_\zeta u)^2}{1-u^2}
 +\frac{(\partial_\zeta v)^2}{1-v^2}\right].
 \label{eq:practicalFI}
\end{equation}
Figure~\ref{fig:conditionalsensitivity} plots the worst-case conditional sensitivity $F_{XY}/T^2$ minimized over the six cardinal Pauli eigenstates of the logical qubit. To match Fig. 3 of the main text, all curves are evaluated at $h = 2$ and a constant dimensionless impulse $\zeta = \lambda T = 0.2$ (so that $\lambda$ scales as $0.2/T$ across the duration scan); Table~\ref{tab:devicebench} reports the numerical values at $T = 20\ \mu\text{s}$ alongside controlled noise ablations.

% When the stored logical state is unknown and repeated across identical sensing runs, state-dependent readout coupling introduces a nuisance parameter into the estimation problem.
When identical copies of an unknown logical state are freshly prepared in repeated sensing runs, state-dependent readout coupling introduces nuisance parameters into the estimation problem.
We use the Fisher-information Schur complement to quantify the information remaining after accounting for an unknown nuisance parameter~\cite{suzuki2020nuisance}.
Consider the one-parameter logical state family $\rho_L(z) = (I + z Z_L)/2$ with unknown coordinate $z \in [-1, 1]$.
Under equal $X/Y$ probe sampling, the two-parameter Fisher information matrix $F_{ab}$ ($a, b \in \{\zeta, z\}$) and the resulting profiled Fisher information $F_{\zeta|z}$ for $\zeta$ are given by
\begin{equation}
 \mathsf F_{ab}=\frac12\left[\frac{\partial_a u\,\partial_b u}{1-u^2}
 +\frac{\partial_a v\,\partial_b v}{1-v^2}\right],\quad
 \mathsf F_{\zeta|z}=\mathsf F_{\zeta\zeta}-\mathsf F_{\zeta z}\mathsf F_{zz}^{+}\mathsf F_{z\zeta}
 =\mathsf F_{\zeta\zeta}(1-\cos^2\theta_F),\quad a,b\in\{\zeta,z\}.
 \label{eq:nuisanceFI}
\end{equation}
% The last form applies for $\mathsf F_{zz}>0$, with $+$ the Moore--Penrose inverse and $|\cos\theta_F|=|\mathsf F_{\zeta z}|/\sqrt{\mathsf F_{\zeta\zeta}\mathsf F_{zz}}$. It is the angle between coherence tangents weighted by the binary-outcome variances. At $\rho_L=I/2$, $T=20\,\mu$s in the stress case, four and six packets give $(\mathsf F_{\zeta\zeta},\mathsf F_{\zeta|z})=(6.0564,0.17209)$ and $(7.80294,0.0014262)$, with $|\cos\theta_F|=0.985690,0.999909$. Even weak logical contamination can therefore produce a large asymptotic penalty through tangent alignment. This model assumes one unknown nuisance component and identically prepared inputs across runs; it does not cover an arbitrary unknown Bloch vector. Exact loss-model independence gives $\partial_z\ell=0$ and no penalty.
Here $F_{zz}^+$ denotes the Moore-Penrose pseudoinverse and $|\cos\theta_F| = |F_{\zeta z}|/\sqrt{F_{\zeta\zeta}F_{zz}}$ (for $F_{zz} > 0$) measures the variance-weighted collinearity between the signal tangent $\partial_\zeta l$
and the logical-nuisance tangent $\partial_z l$ in the complex coherence plane.
Evaluating the combined stress model at $\rho_L = I/2$ and $T = 20\ \mu\text{s}$ yields
$(F_{\zeta\zeta}, F_{\zeta|z}) = (6.0564, 0.17209)$ with $|\cos\theta_F| = 0.985690$ for the four-packet code, and
$(7.80294, 0.0014262)$ with $|\cos\theta_F| = 0.999909$ for the six-packet code.
Thus, whenever the nuisance perturbation $\partial_z l$ aligns nearly parallel to the signal trajectory $\partial_\zeta l$,
even an extremely weak logical coupling can induce a severe asymptotic penalty in $F_{\zeta|z}$.
Note that this model treats a single unknown Bloch coordinate across identically prepared runs rather than an arbitrary three-dimensional Bloch vector;
in the pure-loss limit, exact state independence gives $\partial_z l = 0$ and eliminates this penalty entirely.

% A local estimate formed by projecting the coherence shift onto the signal tangent has linearized bias
In practical finite-shot experiments, however, a bounded coupling $\delta_{\mathrm{read}}$ only induces a tiny systematic shift rather than a catastrophic loss of precision.
Specifically, a standard local estimator that projects the measured coherence shift onto the unperturbed signal tangent $\partial_\zeta l$ incurs a linearized systematic bias of at most
\begin{equation}
 \Delta\zeta(z)=z\frac{\Re[(\partial_z\ell)(\partial_\zeta\ell)^*]}{|\partial_\zeta\ell|^2},\qquad
 |\Delta\zeta|\leq\frac{\delta_{\rm read}}{|\partial_\zeta\ell|}\quad(|z|\leq1).
 \label{eq:nuisancebias}
\end{equation}
% At $T = 20\ \mu\text{s}$ under combined stress, the linearized biases per unit $z$ are merely 
% $3.070 \times 10^{-5}$ and $2.037 \times 10^{-5}$ for the four- and six-packet codes, respectively.
% For the four-packet code, this systematic shift remains smaller than the statistical uncertainty $1/\sqrt{N_{\mathrm{shot}} F_{\zeta\zeta}}$
% up to $N_{\mathrm{shot}} \simeq 1.75 \times 10^8$ shots.
% Bounded finite-sample bias and asymptotic profiled Fisher information thus probe distinct operational regimes.
% Finally, we note that when $\partial_z l$ approaches the numerical truncation floor (such as after ideal Kerr reversal),
% finite-precision tangent angles cannot be used to certify asymptotic profiled sensitivity,
% and adding extra measurement settings removes collinearity only if the resulting nuisance tangents become linearly independent. 
At $T = 20\ \mu\text{s}$ under the combined stress model, the calculated linearized biases per unit $z$ are $3.070 \times 10^{-5}$ and $2.037 \times 10^{-5}$ for the four- and six-packet codes, respectively.
For the four-packet code, this local systematic shift remains smaller than the fixed-state statistical standard deviation
$1/\sqrt{N_{\text{shot}} F_{\zeta\zeta}}$ up to $N_{\text{shot}} \simeq 1.75 \times 10^8$ shots.
We emphasize that this linearized local bias-variance comparison and the asymptotic profiled Fisher information $F_{\zeta|z}$ characterize distinct estimation tasks:
(i) $\min F_{XY}/T^2$ in Fig.~\ref{fig:conditionalsensitivity} evaluates conditional sensitivity for known cardinal Pauli inputs;
(ii) $F_{\zeta|z}$ quantifies asymptotic joint identifiability when an unknown one-parameter state $\rho_L(z)$ is freshly re-prepared in each shot;
(iii) a general unknown three-dimensional logical Bloch vector requires additional measurement settings or structural constraints to ensure joint parameter identifiability;
and (iv) the multi-cycle protocol of Sec.~\ref{App:Repeated} preserves a single unknown quantum state across consecutive sensing windows rather than sampling freshly prepared copies.
Finally, when $\partial_z l$ approaches the numerical truncation floor (as under ideal Kerr reversal), finite-precision tangent angles cannot certify asymptotic profiled sensitivity.   

\subsection{Endpoint control and thermal-event budgets}
Implementing the conditional readout kick $U_h = e^{-i q_{\eta,h} x Z_Q}$ via echoed conditional displacement (ECD) gates~\cite{eickbusch}
requires a phase-space branch separation of $|\beta| = \sqrt{2}q_{\eta,h}$.
Under the dispersive coupling convention $H_{\mathrm{disp}} = \chi n Z_Q / 2$,
an intermediate cavity displacement of amplitude $|\alpha_0|$ achieves this kick in an interaction time
$t_{\mathrm{int}} \simeq |\beta|/(\chi|\alpha_0|)$.
At $T = 20\ \mu\text{s}$ and $h = 2$, we have $q_{\eta,2} \simeq 2.095$ and $|\beta| \simeq 2.963$;
using $|\alpha_0| \in [5, 10]$ with $\chi/2\pi = 42\text{ kHz}$ yields a compact interaction time of
$t_{\mathrm{int}} \simeq 1.12\text{--}2.25\ \mu\text{s}$, excluding finite-pulse shaping overheads and transient photon excursions.
From the reported transmon relaxation time $T_1^q = 110\ \mu\text{s}$ and Hahn-echo coherence time $T_{2,\mathrm{echo}}^q = 80\ \mu\text{s}$,
we extract a pure echo dephasing time of $T_{\phi,\mathrm{echo}}^q = [1/T_{2,\mathrm{echo}}^q - 1/(2T_1^q)]^{-1} = 125.7\ \mu\text{s}$.
Modeling transmon decoherence during the kick as Markovian pure dephasing over $1.1\text{--}2.2\ \mu\text{s}$ predicts a modest Ramsey visibility reduction of $0.87\%\text{--}1.73\%$.  
% For echoed conditional displacement~\cite{eickbusch}, $U_h$ requires branch separation $|\beta|=\sqrt2q_{\eta,h}$. In the convention $H_{\rm disp}/\hbar=\chi nZ_Q/2$, an intermediate displacement $\alpha_0$ gives $t_{\rm int}\simeq|\beta|/(\chi|\alpha_0|)$. At $20\,\mu$s and $h=2$, $q_{\eta,2}\simeq2.095$, $|\beta|\simeq2.963$, and $|\alpha_0|=5$--$10$ gives $t_{\rm int}\simeq2.25$--$1.12\,\mu$s. Finite pulses add overhead and temporary energy. The measured $T_1^q=110\,\mu$s and echo $T_2^q=80\,\mu$s imply $T_{\phi,\rm echo}^q=[1/T_{2,\rm echo}^q-1/(2T_1^q)]^{-1}=125.7\,\mu$s. Extrapolating it as Markovian pure dephasing gives $0.87$--$1.73\%$ visibility loss over the rounded $1.1$--$2.2\,\mu$s range, rather than a gate-fidelity prediction.

Crucially, pure transmon dephasing ($Z_Q$ noise) commutes with an ideal longitudinal kick Hamiltonian
$f(t)x Z_Q$. Denoting the probe dephasing factor by $\nu = e^{-t_{\mathrm{int}}/T_\phi^q}$
and the corresponding dephasing channel by
$\mathcal{D}_\nu^Q(\sigma) = [(1 + \nu)\sigma + (1 - \nu)Z_Q \sigma Z_Q]/2$, the noisy kick channel obeys
\begin{equation}
 \mathcal E_{\rm kick,\phi}=(\id\otimes\mathcal D_\nu^Q)\circ\mathcal U_h,\qquad
 \Tr_Q\mathcal E_{\rm kick,\phi}(\rho)=\Tr_Q\mathcal U_h(\rho).
 \label{eq:probedephasing}
\end{equation}
% It reduces visibility but adds no unconditional memory error after fixed probe-independent recovery, including with a reference. Conditional records, transmon relaxation, and finite transverse pulses require separate analysis. Assume that a ground-state-reset transmon undergoes state-independent Poisson excitation with $T_\uparrow^q\simeq92$ ms, and that its no-event branch follows the modeled baseline channel. Then $p_\uparrow=1-e^{-T/T_\uparrow^q}\simeq2.17\times10^{-4}$ at $20\,\mu$s bounds the additional entanglement infidelity and half diamond distance when all event trajectories are treated as arbitrary errors. Preparation, extraction, recovery, and initial thermal population still require their own budgets.
Consequently, while pure probe dephasing reduces sensing contrast by $\nu$,
it introduces zero unconditional backaction onto the reduced quantum memory after probe-independent recovery, including for reference-entangled inputs.
Conditioning on individual probe measurement outcomes, transmon $T_1^q$ relaxation during the gate,
and finite-bandwidth transverse control pulses break this exact commutation and require dedicated pulse-level compilation.
During the sensing interval $T$ itself, the ancillary transmon remains parked in its ground state $|g\rangle$;
given the reported thermal heating time $T_\uparrow^q \simeq 92\text{ ms}$ \cite{milul},
the probability of a parasitic thermal excitation over $T = 20\ \mu\text{s}$ is $p_\uparrow = 1 - e^{-T/T_\uparrow^q} \simeq 2.17 \times 10^{-4}$.
Treating any thermal jump trajectory as a worst-case arbitrary error bounds the resulting increment in both memory entanglement infidelity and half diamond-norm distance by $p_\uparrow$.
State preparation, syndrome extraction, recovery gates, and residual cavity thermal population ($< 0.5\%$) similarly require separate fault-tolerant error budgets.  

\subsection{Repeated sensing with one stored logical state}\label{App:Repeated}
% Coherent re-encoding transfers the recovered unknown qubit into a fresh common packet, with a fresh probe and independent environmental increments. For fixed-duration cycles sensing one constant $\lambda$, the product approximation $\mathcal D_p\otimes\rho_Q(\lambda)$ gives
Because our protocol actively recovers the encoded logical state at the end of each sensing window,
an unknown quantum state can be preserved across $m$ consecutive sensing cycles by coherently re-encoding the recovered logical register into a fresh wave-packet state while resetting the probe qubit.
For $m$ identical cycles of duration $T$ monitoring a quasi-static force $\lambda$ under independent Markovian loss increments,
the leading factorized channel approximation $\mathcal{D}_p \otimes \rho_Q(\lambda)$ yields the cumulative memory infidelity and total quantum Fisher information

\begin{equation}
 1-F_e^{(m)}=\frac{1-(1-2p)^m}{2}=mp+O(m^2p^2),\qquad F_Q^{(m)}=mF_Q.
 \label{eq:repeatedbudget}
\end{equation}
% For four packets with $L=3$, $w=1.8$, $\gamma=10^{-3}$, and $h=2$, the exact reduced original-label calculation gives $p\simeq1.158682\times10^{-5}$ and $F_Q/T^2=9.12958$.
For the four-packet code at $L = 3$, $w = 1.8$, $\gamma = 10^{-3}$, and $h = 2$,
exact original-label evaluation gives a per-cycle dephasing probability $p \simeq 1.158682 \times 10^{-5}$
and a normalized single-cycle sensitivity $F_Q/T^2 = 9.12958$.

Within this factorized product approximation, a single stored qubit can sustain up to $m = 86$ consecutive sensing cycles
while keeping the cumulative memory infidelity below a $10^{-3}$ threshold ($1 - F_e^{(86)} \simeq 9.955 \times 10^{-4}$, whereas $87$ cycles exceed $10^{-3}$),
achieving an ideal multi-cycle Cram\'{e}r-Rao sensitivity scale of $(\sqrt{m F_Q})^{-1} \simeq 0.0357/T$ for $\lambda$.
Beyond the leading product approximation, if each physical cycle deviates from the factorized target by at most $\epsilon$ in the half diamond norm,
a standard telescoping-sum argument bounds the $m$-cycle cumulative channel discrepancy by $\min(1, m\epsilon)$ (treating flagged erasures as an absorbing subspace).
A complete experimental error budget must combine this accumulated channel residual with finite-duration re-encoding and recovery gate errors,
while the effective metrological bandwidth scales with the total cycle time $T_{\mathrm{cycle}} = T + t_{\mathrm{overhead}}$.
Notably, this multi-cycle protocol preserves a single unknown quantum state throughout the entire measurement record, in contrast to the nuisance-parameter analysis of Sec.~\ref{sec:sensitivity},
which assumes freshly re-prepared copies of an unknown state in each shot.

\begin{inlinetable}
\caption{Superconducting cavity-transmon hardware parameters adopted from Ref.~\cite{milul} and their roles in our benchmark analysis.
The intrinsic cavity dephasing time $T_\phi = 0.5\text{ s}$ represents a conservative stress assumption corresponding to the reported experimental lower bound,
and the residual Kerr coefficient $K$ reflects the simulated value reported in Ref.~\cite{milul}.
}
\label{tab:device}
\begin{ruledtabular}
\begin{tabular}{lll}
Quantity&Reported value&Role\\
Cavity $T_1^c$&25.6 ms&Photon-loss rate\\
Cavity $T_2^c$&34 ms&Includes thermal-transmon events\\
$K/2\pi$&3.6 Hz&Residual Kerr\\
Intrinsic cavity $T_\phi$&$>0.5$ s&Stress case: set to 0.5 s\\
$\chi/2\pi$&42 kHz&Endpoint interaction estimate\\
Transmon $T_1^q$, $T_2^q$, $T_{2,\rm echo}^q$&110, 16, 80 $\mu$s&Control context\\
Transmon $T_\uparrow^q$&$\simeq92$ ms&Separate event budget\\
Cavity thermal population&$<0.5\%$&Separate gain budget
\end{tabular}
\end{ruledtabular}
\end{inlinetable}

\section{Numerical details}
\label{sec:numerics}

For the pure-loss physical-channel benchmarks, we evaluate the position-space integral in Eq.~\eqref{eq:poskernel}
using Gauss--Hermite quadrature for the environment variable $\xi$
combined with finite-cell Gauss--Legendre quadrature for the spatial trace in Eq.~\eqref{eq:extract},
retaining the entangled reference, probe qubit, and erasure flag throughout. Normalized Choi trace distances $D_{\mathrm{Choi}}$
and factorization residuals ($D_{\mathrm{corr}}, D_{\mathrm{fact}}$) are computed directly from these exact physical channels as defined in the End Matter.
Table~\ref{tab:hierarchy} summarizes the numerical loss hierarchy for four through ten wave packets:
the fitted scaling exponents for $1 - F_e$ and $2D_{\mathrm{Choi}}$ verify the analytical leading-order memory dephasing of Eq.~\eqref{concise:hierarchy},
while the four-packet Choi residuals isolate the third-order memory-probe decoupling.

\begin{inlinetable}
\caption{
    Numerical pure-loss hierarchy for the physical channel at $L = 3$, $w = 1.8$, $\lambda T = 0.2$, and $h = 1$. Here $\bar{n}$ is the initial mean photon number,
and $a_r$ is the analytical leading-order dephasing prefactor in Eq.~\eqref{concise:hierarchy}.
Values of $1 - F_e$ and $2D_{\mathrm{Choi}}$ are reported at $\gamma = 10^{-3}$, alongside log-log scaling exponents fitted over the resolved low-loss window. 
    % Physical-channel loss hierarchy at $L=3$, $w=1.8$, $\lambda T=0.2$, and $h=1$. $a_r$ is the coefficient of $\gamma^{r+1}$ in Eq.~\eqref{concise:hierarchy}. Both exponents are finite-window fits; errors and $2D_{\rm Choi}$ are evaluated at $\gamma=10^{-3}$.
    }
\label{tab:hierarchy}
\begin{ruledtabular}
\begin{tabular}{cccccccc}
Packets&Packet power&$\bar n$&$a_r$&$1-F_e$&Exponent&$2D_{\rm Choi}$&Joint exponent\\
4&4&6.37451&11.390625&$1.15868\times10^{-5}$&2.013&$2.317\times10^{-5}$&2.013\\
6&6&10.12225&213.574219&$2.21116\times10^{-7}$&3.026&$4.422\times10^{-7}$&3.026\\
8&8&13.88315&5886.639404&$6.2381\times10^{-9}$&4.040&$1.248\times10^{-8}$&4.040\\
10&10&17.64906&214568.006287&$2.34\times10^{-10}$&5.08&$4.681\times10^{-10}$&5.081
\end{tabular}
\end{ruledtabular}
\end{inlinetable}

% Fits use $\gamma=0.0002,0.0005,0.001,0.002$ for four through eight packets, and omit $0.0002$ for ten packets because its memory error lies below the empirical floor $6.93\times10^{-12}$. Increasing the quadratures from $(n_y,n_h)=(320,100)$ to $(480,160)$ changes Choi entries by at most $1.78\times10^{-13}$; an independent 70-digit original-label calculation at $\gamma=10^{-5}$ gives error-to-leading-term ratios $1.00017,1.00036,1.00060,1.00091$. An independent Fock readout check at cutoff 240 has coherence discrepancy below $7\times10^{-9}$.

The power-law exponents in Table S2 are obtained from log-log fits over
$\gamma \in \{0.0002, 0.0005, 0.001, 0.002\}$ for four, six, and eight packets,
whereas the ten-packet fit omits $\gamma = 0.0002$ because its memory infidelity falls below our empirical double-precision numerical floor ($6.93 \times 10^{-12}$).
Increasing the quadrature resolution from $(n_y, n_h) = (320, 100)$ to $(480, 160)$ alters individual Choi-matrix entries by at most $1.78 \times 10^{-13}$.
Furthermore, an independent 70-digit arbitrary-precision evaluation of the virtual original-label channel at $\gamma = 10^{-5}$ yields numerical-to-analytical
leading-term ratios of $1.00017, 1.00036, 1.00060,$ and $1.00091$ for four through ten packets, respectively.
Cross-checking the probe readout against an independent Fock-space propagation truncated at $N_{\mathrm{cut}} = 240$ confirms agreement in probe coherence to within $7 \times 10^{-9}$.

For the four-packet factorization residuals evaluated over
$\gamma \in [5 \times 10^{-4}, 4 \times 10^{-3}]$ at $\lambda T = 0.2$, as shown in Fig. 4(b) of the main text,
log-log fits yield cubic scaling exponents of $3.000046$ ($h = 1$) and $2.993608$ ($h = 2$) for $D_{\mathrm{corr}}$, and $2.987250$ (at both $h = 1, 2$)for $D_{\mathrm{fact}}$.
Refining the quadrature grid up to $(n_y, n_h) = (640, 224)$ changes the computed Choi trace distances by less than $6.89 \times 10^{-14}$,
and an independent 60-digit Fourier-kernel evaluation agrees to within $7.75 \times 10^{-16}$.
The smallest plotted residual remains a factor of $59.8$ above the empirical zero-loss numerical floor ($1.378 \times 10^{-11}$).
We note that these quadrature refinement checks serve as convergence diagnostics rather than certified interval-arithmetic bounds or exact semidefinite-programming evaluations of the diamond norm.

To simulate the master Eq. \eqref{eq:deviceME} for the superconducting-cavity benchmarks,
we project the continuous-variable encoding states onto a truncated Fock basis to form the matrix $A$,
and apply symmetric orthonormalization $V_F = A(A^\dagger A)^{-1/2}$ to preserve exact isometry on the discrete grid.
Propagating the full four-element logical operator basis retains complete channel coherence and reference entanglement.
The numerical implementation combines QuTiP~\cite{qutip2013,qutip2026} for Liouvillian construction and master-equation propagation with SciPy sparse-exponential routines for signal derivatives.
Signal derivatives with respect to $\zeta = \lambda T$ are computed accurately via the Fr\'{e}chet derivative of the Liouvillian exponential using a block-triangular auxiliary matrix,
and are independently validated against centered finite differences.
Increasing the Fock cutoff and spatial decoder quadrature from
$(N_{\mathrm{cut}}, n_y) = (320, 360)$ to $(400, 440)$ across 16 benchmark configurations at $T = 20$ and $80\ \mu\text{s}$ changes individual Choi matrix elements,
memory infidelities $1 - F_e$, and worst-case sensitivities $\min F_{XY}/T^2$ by at most $8.23 \times 10^{-10}$, $1.66 \times 10^{-9}$, and $7.54 \times 10^{-8}$, respectively.
At $\gamma = 10^{-3}$, the second-harmonic readout ($h = 2$) yields analytical sensitivities
$F_Q/T^2 = 9.12958, 11.15419, 12.39118,$ and $13.22184$ for four through ten packets, compared to $3.732270$ for the four-packet fundamental harmonic $h = 1$,
while physical-cell simulations at $\gamma = 0.001$ and $0.008$ confirm that the reduced-memory infidelity remains identical between $h = 1$ and $h = 2$ to within numerical precision.

Finally, the controlled ablation results in Table~\ref{tab:devicebench}(b) elucidate the non-monotonic code ordering observed in Fig.~3(b) of the main text.
Setting $K = 0$ (and even isolating pure photon-number dephasing $\mathcal{D}[n]$ alone) reproduces
both the superiority of the six-packet code over the eight-packet code and the crossover
where the six-packet code overtakes the ten-packet code at $T = 80\ \mu\text{s}$.
This confirms that the post-Kerr-reversal performance hierarchy is governed by code-dependent higher-order number-dephasing moments ($O((T/T_\phi)^2)$),
which grow rapidly with photon-number variance, rather than by residual noncommuting Kerr cross terms.
While our first-order analysis in Sec.~\ref{sec:dephasing} proves that all codes with $r \ge 2$ eliminate $O(T/T_\phi)$ dephasing errors,
it does not constrain the prefactors of these second-order dephasing terms.
All ablation benchmarks were evaluated at $(N_{\mathrm{cut}}, n_y) = (400, 480)$ and cross-checked against $(320, 400)$,
confirming convergence of $1 - F_e$ to within $8.12 \times 10^{-12}$ above a zero-noise floor below $5.84 \times 10^{-13}$.
% Across all numerical channel evaluations, trace preservation, Hermiticity, and complete positivity were verified to machine precision.

\begin{inlinetable}
\caption{
    Superconducting-cavity benchmarks and controlled noise ablations at $L = 3$, $w = 1.8$, $h = 2$, and $\lambda T = 0.2$,
    assuming ideal endpoint controls and excluding thermal transmon jumps.
    (a) Comparison at $T = 20\ \mu\text{s}$ between pure photon loss, the combined stress model (loss, Kerr, and $T_\phi = 0.5\text{ s}$),
    and the stress model followed by ideal endpoint Kerr reversal; $\mathrm{FI}$ denotes $\min F_{XY}/T^2$.
    (b) Controlled ablation study comparing memory entanglement infidelity $1 - F_e$ after ideal Kerr reversal against an identical channel with $K = 0$,
    as well as pure photon-number dephasing alone ($\kappa = \lambda = K = 0$).
    % Device benchmarks and controlled noise ablations, with $L=3$, $w=1.8$, $h=2$, and ideal endpoint operations. (a) At $T=20\,\mu$s, stress includes loss, Kerr, and $T_\phi=0.5$ s; FI means $\min_6F_{XY}/T^2$. (b) Memory errors after ideal inverse Kerr are compared with $K=0$ at the same loss and $\lambda T=0.2$, and with pure number dephasing alone (no loss, force, or Kerr). All outcomes are retained; thermal transmon events are excluded.
    }
\label{tab:devicebench}
\begin{ruledtabular}
\begin{tabular}{cccccc}
\multicolumn{6}{l}{(a) Common-harmonic comparison at $20\,\mu$s}\\
&Loss only&\multicolumn{2}{c}{Stress}&\multicolumn{2}{c}{Stress + inverse Kerr}\\
Packets&$1-F_e$&$1-F_e$&FI&$1-F_e$&FI\\
4&$7.046\times10^{-6}$&$1.488\times10^{-3}$&6.0561&$1.216\times10^{-3}$&6.0955\\
6&$1.047\times10^{-7}$&$3.813\times10^{-4}$&7.8028&$5.657\times10^{-5}$&8.0153\\
8&$2.295\times10^{-9}$&$2.700\times10^{-3}$&8.7226&$1.484\times10^{-4}$&9.3179\\
10&$6.69\times10^{-11}$&$3.080\times10^{-3}$&9.0358&$2.865\times10^{-5}$&10.2410
\end{tabular}

\medskip
\begin{tabular}{ccccc}
\multicolumn{5}{l}{(b) Controlled ablations: memory entanglement infidelity}\\
$T\ (\mu\mathrm{s})$&Packets&Stress + inverse Kerr&$K=0$&Pure $\mathcal D[n]$\\
20&6&$5.657\times10^{-5}$&$5.440\times10^{-5}$&$2.236\times10^{-5}$\\
20&8&$1.484\times10^{-4}$&$1.477\times10^{-4}$&$1.480\times10^{-4}$\\
20&10&$2.865\times10^{-5}$&$2.740\times10^{-5}$&$1.671\times10^{-5}$\\
80&6&$8.837\times10^{-4}$&$8.461\times10^{-4}$&$3.449\times10^{-4}$\\
80&8&$2.098\times10^{-3}$&$2.061\times10^{-3}$&$2.070\times10^{-3}$\\
80&10&$1.274\times10^{-3}$&$1.190\times10^{-3}$&$7.407\times10^{-4}$
\end{tabular}
\end{ruledtabular}
\end{inlinetable}

% \bibliography{references}
%apsrev4-2.bst 2019-01-14 (MD) hand-edited version of apsrev4-1.bst
%Control: key (0)
%Control: author (8) initials jnrlst
%Control: editor formatted (1) identically to author
%Control: production of article title (0) allowed
%Control: page (0) single
%Control: year (1) truncated
%Control: production of eprint (0) enabled
%